\documentclass[entropy,preprints,article,accept,moreauthors,pdftex]{mdpisalvo2} 

\usepackage{xcolor}

\newcommand\aap{Astron. Astrophys.}                
\newcommand\apj{Astrophys. J.}                 
\newcommand\apjl{Astrophys. J. Lett.}                
\newcommand\apjs{Astrophys. J. Suppl. Ser.}               

\newcommand\araa{Annu. Rev. Astron. Astrophys.}             
\newcommand\mnras{Mon. Not. R. Astron. Soc.}             
\newcommand\nat{Nature}              
\newcommand\pra{Phys. Rev.~A}        
\newcommand\solphys{Sol.~Phys.}      

\newcommand\arcsec{\mbox{$^{\prime\prime}$}}%
\usepackage{amssymb}
\usepackage{amsthm,amsmath}
\usepackage[normalem]{ulem} 
\usepackage{soul} 
\usepackage{natbib}
\firstpage{1} 
\pubvolume{1}
\issuenum{1}
\articlenumber{0}
\pubyear{2021}
\copyrightyear{2021}
\externaleditor{Academic Editor: José A. Tenreiro Machado } 
\datereceived{24 February 2021} 
\dateaccepted{28 March 2021} 
\datepublished{} 
\hreflink{https://doi.org/} 

\Title{Analysis of Pseudo-Lyapunov Exponents of Solar Convection Using State-of-the-Art Observations}

\TitleCitation{Analysis of Pseudo-Lyapunov Exponents of Solar Convection Using State-of-the-Art Observations}

\Author{Giorgio Viavattene 
 $^{1,}$*\orcidA{}, Mariarita Murabito $^{1}$\orcidB{}, Salvatore L. 
 Guglielmino $^{2}$\orcidC{}, Ilaria Ermolli $^{1}$\orcidD{}, \mbox{Giuseppe Consolini $^{3}$\orcidE{}}, Fabrizio Giorgi $^{1}$ and Shahin Jafarzadeh $^{4,5,}$\orcidF{}}

\AuthorNames{Giorgio Viavattene, Mariarita Murabito, Salvatore L. Guglielmino, Ilaria Ermolli, Giuseppe Consolini, Fabrizio Giorgi and Shahin Jafarzadeh}

\AuthorCitation{Viavattene, G.; Murabito, M.; Guglielmino, S.L.; Ermolli, I.; Consolini, G.; Giorgi, F.; Jafarzadeh, S.}

\address{%
$^{1}$ \quad INAF---Osservatorio Astronomico di Roma, Via Frascati 33, I-00078 Monte Porzio Catone, Italy; mariarita.murabito@inaf.it (M.M.); ilaria.ermolli@inaf.it (I.E.); fabrizio.giorgi@inaf.it (F.G.)\\
$^{2}$ \quad INAF---Osservatorio Astrofisico di Catania, Via S. Sofia 78, I-95123 Catania, Italy; salvatore.guglielmino@inaf.it\\
$^{3}$ \quad INAF---Istituto di Astrofisica e Planetologia Spaziali, Via del Fosso del Cavaliere 100, I-00133 
Rome, Italy; giuseppe.consolini@inaf.it\\
$^{4}$ \quad Rosseland Centre for Solar Physics, University of Oslo, P.O. Box 1029 Blindern, NO-0315 Oslo, Norway\\
$^{5}$ \quad Institute of Theoretical Astrophysics, University of Oslo, P.O. Box 1029 Blindern, NO-0315 Oslo, Norway}

\corres{Correspondence: giorgio.viavattene@inaf.it (G.V.); \mbox{Tel.: +39-06-94286-471 (G.V.)}}

\abstract{The solar photosphere and the outer layer of the Sun's interior are characterized by convective motions, which display a chaotic and turbulent character. In this work, we evaluated the pseudo-Lyapunov exponents of the overshooting convective motions observed on the Sun's surface by using a method employed in the literature to estimate those exponents, as well as another technique deduced from their definition.~We analyzed observations taken with state-of-the-art instruments at ground- and space-based telescopes, and we particularly benefited from the spectro-polarimetric data acquired with the Interferometric Bidimensional Spectrometer, the Crisp Imaging SpectroPolarimeter, and the Helioseismic and Magnetic Imager. Following previous studies in the literature, we computed maps of four quantities which were representative of the physical properties of solar plasma in  each observation, and estimated the pseudo-Lyapunov exponents from the residuals between the values of the quantities computed at any point in the map and the mean of values over the whole map. In contrast to previous results reported in the literature, we found that the computed exponents hold negative values, which are typical of a dissipative regime, for all the quantities derived from our observations. The values of the estimated exponents increase with the spatial resolution of the data and are almost unaffected by small concentrations of magnetic field. Finally, we showed that similar results were also achieved by estimating the exponents from residuals between the values at each point in maps derived from observations taken at different times. The latter estimation technique better accounts for the definition of these exponents than the method employed in previous studies.}


\begin{document}
\section{Introduction}\label{sec1}

Turbulent thermal convection is a non-equilibrium process ubiquitous in nature that sets in at a high Rayleigh number (Ra) \citep{niemela2000}. It occurs in several astrophysical environments, such as stellar interiors and planetary atmospheres, and it owns some aspects that are not completely understood. A typical manifestation of this process in the heliosphere is represented by the solar convection \citep{nordlund2009}.

In the outermost layers of the Sun's interior ($r \gtrsim 0.7 \,R_{\odot}$), the energy generated by nuclear reactions is transported almost entirely by convection, because radiative transfer becomes inefficient owing to the increase in opacity determined by the recombination of hydrogen and helium \citep{nordlund2009}.~The convective envelope extends up to the solar surface, where plasma becomes optically thin and photons with energy in the visible range are eventually radiated into space. At a small scale, the region radiating the solar energy, called the photosphere, appears to be permeated by inhomogeneities that give rise to a granular pattern, also known as solar granulation \citep{nordlund2009}. The granules, characterized by a spatial dimension of 1--2 Mm and a typical lifetime of 5--10 min, are due to the plasma motions that are driven by the convective overshooting, thus reflecting the turbulent motions underneath the solar surface, where Ra is estimated to be of the order of $10^{23}$~\citep{niemela2000}. 
A photospheric pattern attributable to the granulation was already observed in the early 19th century \citep{herschel1801}. However, it was only in recent decades that our knowledge of solar convection has greatly advanced. Indeed, relatively long series of homogeneous observations are needed to disclose the attributes and various manifestations of solar convection. These series have been achieved either in space or by equipping ground-based telescopes with adaptive optics (AO) systems that compensate the blurring and distortion in the observation due to the Earth's atmospheric turbulence (seeing). Current 1 m class telescopes operating with such compensation allow acquiring the time series of data of the solar surface that last several tens of minutes by resolving spatial scales down to 100~km. Recently, the first 4 m class solar telescope, the Daniel K.~Inouye Solar Telescope (DKIST, \citep{rimmele2019}) in Haleakala (Hawaii, USA) has acquired data of the photosphere resolving even smaller spatial scales~\citep{johnston2020}. Furthermore, the residual degradation of these data originating from the instrument, due to optical distortions and light scattering, and unaccounted for by the AO systems, can now be removed from the observations with the application of post facto image processing techniques \citep{lofdahl2002}.

The imaging observations of the photosphere obtained with these methods have provided information about the morphology and evolution of the granulation pattern \citep{hirzberger1997,hirzberger1999}, its dynamics including vortex flows \citep{lemmerer2017} and advection by larger scale flows \citep{muller1992}. 
Moreover, the analysis of spectral lines measurements, which reveals the properties of the convective plasma logged in the line shape, has shown the height-dependence of the velocity structure of flows in granules and intergranular lanes \citep{oba2017a,oba2017b} and the narrow transition zones where granular flows bend down \citep{khomenko2010}, as well as the presence of the small-scale turbulent motions therein \citep{ishikawa2020}. 
Some studies have also reported convective patterns at scales larger than those of the granulation, in particular of the mesogranulation (with a spatial scale of 5--10 Mm, e.g., \citep{november1981,brandt1991,cerdena2003,leitzinger2005,rast2003}), supergranulation (with a spatial scale of 20--50 Mm, \citep{berrilli2004,delmoro2004,rincon2018}) and giant cells (with a spatial scale of 100 Mm and larger, \citep{beck1998,hathaway2013,mcintosh2014}). However, the real presence of some of these larger scale convective patterns is still debated \citep{rieutord2000,rast2003,rincon2018}. 

High-resolution observations have also shown that the magnetic field emerging into the solar atmosphere strongly affects the granulation. Generated by a turbulent dynamo process in the Sun's interior, the magnetic field is maintained by converting the kinetic energy of convective plasma motions into magnetic energy \citep{charbonneau2020}. Therefore, there is an interplay between the magnetic field and plasma flows, which at small scales also manifest as the so-called quiet-Sun magnetism (see \citep{stein2012,bellotrubio2019} for a review). The interaction between granules and magnetic fields is not completely clear and there are several scientific open questions on the involved processes, especially related to the emergence of magnetic flux concentrations at the granular spatial scale \citep{centeno2007,orozcosuarez2008,viticchie2012} and their subsequent diffusion \citep{giannattasio2013,giannattasio2014,jafarzadeh2017}. In this context, some authors have highlighted the presence of ubiquitous horizontal fields in the internetwork \citep{lites1996,lites2008,danilovic2010,kianfar2018}. Other authors have studied the evolution of coherent magnetic flux structures emerging in the quiet Sun. For example, the emergence of magnetic bipoles in the quiet Sun has been analyzed \citep{martinez2009}; the observation of a small-scale magnetic bipole has been reported from emergence to decay, studying the correlation between its evolution and the underlying convective motions \citep{guglielmino2012}; flux-sheet emergence events consisting of strongly inclined magnetic fields  appearing at the granular scale have recently been analyzed~\citep{Fischer2019,guglielmino2020}; and the difference between the granules observed in various regions characterized by different field strengths has been investigated \citep{falco2017}. 

In addition to solar physics studies, high-resolution observations of the Sun's convection have also been a natural test-bed for the theory of dynamical systems and non-linear dynamics.~\citet{nesis2001} 
have applied non-linear methods to series of spectral observations to investigate the attractor underlying the granulation phenomenon in the three-dimensional phase space spanned by intensity, Doppler velocity, and line broadening. They have found that the granulation attractor does not fill the entire phase space, as one would expect from the high Reynolds and Ra numbers of the photospheric plasma. In addition, using photospheric observations in the quiet Sun, \citet{viavattene2020} have recently shown that solar convective motions satisfy the symmetry relations of the Gallavotti--Cohen fluctuation relation for non-equilibrium stationary states.

It is worth mentioning that multi-dimensional time-dependent atmosphere models derived from radiative magneto--hydrodynamic (MHD) simulations are able to reproduce the solar granulation pattern with great fidelity \citep{gudiksen2011,beeck2012,freytag2012}. This might lead to the conclusion that the atmosphere models derived from simulations are accurate representations of the real Sun \citep{nordlund2009}. However, simulations suffer from limitations and uncertainties on the physical parameters employed and can only cover a part of the size range of the solar convection. 

The current knowledge derived from both observations and simulations still does not allow us to understand all the processes involved with solar convection. For example, the convection also plays a role in the excitation and upwards propagation of waves into the solar atmosphere \citep{goldreich1990}. Indeed, turbulent motions are expected to excite waves that propagate outwards, transporting mechanical energy through the photospheric layers into the chromosphere and corona \citep{carlsson2019}. However, while the occurrence of acoustic events has been demonstrated in the quiet Sun \citep{malherbe2012,malherbe2015}, with waves propagating up to the upper atmospheric layers \citep{kayshap2018}, the excitation of MHD waves by the solar convection, predicted by numerical simulations, has been rarely observed \citep{morton2014,stangalini2014,stangalini2017}.

Here, we focused on the nature of the solar convection, aiming to further characterize it by means of the pseudo-Lyapunov exponents. 

In the framework of chaotic and dynamical systems, Lyapunov exponents (hereafter also referred to as $\Lambda$ exponents) are quantities that characterize the rate of divergence of the trajectories described by the system in the phase space, thus providing information about its chaotic or dissipative state. In other words, if the physical system begins to evolve from two slightly different initial states $x$ and $x+\epsilon$, the divergence between these initial states after $n$ iterations is given by
\begin{equation}
\epsilon(n) \sim \epsilon e^{\Lambda n}
\label{EQ1}
\end{equation}
If $\Lambda > 0$, the trajectories will diverge exponentially and the system is chaotic. Conversely, if $\Lambda < 0$, the system is in dissipative regime.~Therefore, the {$\Lambda$} exponents indicate the predictability for dynamical systems, considered as an important tool for studying the stability of a dynamical system.

The Lyapunov exponents of the solar granulation were estimated in model atmospheres derived from the numerical simulations as in \citet{kurths1991,steffen1995}, by varying the control parameters (e.g., Ra, Prandtl, or Taylor numbers) of computations. In addition, they were derived from the spectral observations of the photosphere by \citet{hanslmeier1994}. The latter study was performed on the photographic spectra of a large sample of granules observed simultaneously with a spatial resolution of $\approx$0.24$\arcsec$, by using a ground-based telescope, which at the time was lacking AO systems.

In view of the lack of time series of observations, \citet{hanslmeier1994} computed the spatial fluctuations of four parameters derived from the available data and estimated the variation of these fluctuations for increasing samples of data points. In fact, \citet{hanslmeier1994} argued that, by virtue of the ergodic condition, the fluctuations of quantities  estimated over a large sample of granules are at some time  representative of  the fluctuations of the same quantities evaluated over a long time interval at a single point of the studied system.~However, \citet{hanslmeier1994} note that the variation in the quantities estimated in their study are not identical to the Lyapunov exponents $\Lambda$  that are inherently representative of the temporal evolution of the studied system. Based on that, \citet{hanslmeier1994} referred to the exponents considered in their study as to Lyapunov-like exponents (hereafter referred to as $\lambda$). The four quantities derived from the observations by \citet{hanslmeier1994} are: the continuum intensity ($I_c$), 
the line-of-sight (LoS) velocity ($v_{LoS}$), the full width at half maximum ($FWHM$) and the asymmetry ($A$) of the spectral lines in their data.

\citet{hanslmeier1994} reported significant changes of the $\lambda$ exponents computed from various line parameters derived from their observations. These changes occurred in a range of spatial scales spanning from the mesogranular to subgranular scale.~In particular, they reported positive values of the $\lambda$ exponents computed from fluctuations of $I_c$ and $A$, when looking at subgranular scales through Fourier filtering. This led the authors to suppose that their measurements were depicting turbulent motions occurring on subgranular spatial scales.

The results obtained by \citet{hanslmeier1994} show a weak convergence and noisy behavior of the computed exponents at the smallest spatial scales investigated by the authors. Moreover, they were derived from a single observation of a sample of granules simultaneously observed  at different evolutionary stages, as a proxy of the temporal evolution of an individual convective granular cell, as for the original definition of the Lyapunov exponents $\Lambda$.~These facts have inspired us to perform a new study of the $\lambda$ exponents based on current state-of-the-art observations. 

In the present work, we analyzed the four spectral line parameters considered by \citet{hanslmeier1994}, and derived them from spectro-polarimetric observations taken with state-of-the-art instruments. The studied data were taken in the quiet Sun near disk center, far away from active regions, at two lines originating in the photosphere and with different spatial resolution. To perform our study, we  employed the methods proposed by \citet{hanslmeier1994}, but we also estimated the exponents from the analysis of the time series of observations as deduced from the definition of the Lyapunov exponents $\Lambda$. However, the Lyapunov exponents computed in our case do not exactly adhere to the same strict definition, so we will refer to these as pseudo-Lyapunov exponents.

In the following sections, after describing the data sets used in this work and the adopted data analysis (Section~\ref{sec:datasets}), we present and discuss the results obtained from our investigation (Sections~\ref{sec:results} and~\ref{sec:final}).

\section{Observations and Data Analysis}
\label{sec:datasets}

\subsection{Observations}

In this study, we analyzed observations of the photosphere obtained with three state-of-the-art instruments which operated at ground-based telescopes with AO systems and in space. The data were taken at the disk center, to prevent projection effects. 
The overall characteristics of each data set are reported in Table~\ref{table1}, which also lists the formation height of the core of the photospheric lines sampled by the analyzed observations. 
\end{paracol}
\nointerlineskip
\begin{specialtable}[H]
\tablesize{\small}
\label{TAB1}
\caption{{Ground- and space-based data sets used for the analysis. References for the formation height of each line are provided.} \label{table1}}
\setlength{\cellWidtha}{\columnwidth/7-2\tabcolsep-0.2in}
\setlength{\cellWidthb}{\columnwidth/7-2\tabcolsep-0.2in}
\setlength{\cellWidthc}{\columnwidth/7-2\tabcolsep-0in}
\setlength{\cellWidthd}{\columnwidth/7-2\tabcolsep+0.3in}
\setlength{\cellWidthe}{\columnwidth/7-2\tabcolsep+0.0in}
\setlength{\cellWidthf}{\columnwidth/7-2\tabcolsep-0.2in}
\setlength{\cellWidthg}{\columnwidth/7-2\tabcolsep+0.3in}
\scalebox{1}[1]{\begin{tabularx}{\columnwidth}{>{\PreserveBackslash\centering}m{\cellWidtha}>{\PreserveBackslash\centering}m{\cellWidthb}>{\PreserveBackslash\centering}m{\cellWidthc}>{\PreserveBackslash\centering}m{\cellWidthd}>{\PreserveBackslash\centering}m{\cellWidthe}>{\PreserveBackslash\centering}m{\cellWidthf}>{\PreserveBackslash\centering}m{\cellWidthg}}
\toprule
\textbf{Telescope} & \textbf{Instrument} & \textbf{Spectral Coverage} & \textbf{Time Coverage} & \textbf{Time Cadence} & \textbf{Spatial Resolution} & \textbf{Formation Height of Line Cores} \\
\midrule
DST &    IBIS     &   Fe I 617.3 nm     &  1 May 2015 14:18--15:03 UT    &  48 s  & 0.16$\arcsec$  & $\sim$150 km \citep{nagashima2014} \\ 
SST &    CRISP    &   Fe I 630.15 nm    &  25 May 2017 09:30--09:40 UT  &  -  & 0.13$\arcsec$  & $\sim$180 km \citep{jafarzadeh2015}  \\ 
SDO &    HMI      &   Fe I 617.3 nm                               &  {1 May 2015 14:24 UT 25 May 2017 09:36 UT}    & - & 1$\arcsec$ & - \\ 
\bottomrule
\end{tabularx}} 

\end{specialtable}

\begin{paracol}{2}
\switchcolumn

The first data set was acquired on 1 May 2015 using the Interferometric Bidimensional Spectrometer (IBIS, \citep{cavallini2006}) which at that time was installed at the Dunn Solar Telescope (DST), located in Sacramento Peak, New Mexico (USA). IBIS acquired the four Stokes~I, Q, U and V parameters with a spectral pass-band of 4~pm along the Fe~I 617.3~nm spectral line with a field-of-view (FoV) of $25\arcsec \times 60\arcsec$. The angular resolution of the data set was $0.16\arcsec$, which corresponds to a spatial resolution of $\simeq$130~km on the solar photosphere. The temporal resolution, which is the time needed to perform a single spectral scan, is 48~s and the whole data set consists of 50~scans. Figure~\ref{FIG1} (top-left panel) shows the FoV of the IBIS data set, where the purple and yellow boxes indicate the two sub-FoVs chosen as representative of magnetic and quiet regions, details of which can be found in the following. The data set was calibrated using the IBIS standard calibration pipeline (e.g.,~\citep{ermolli2017,murabito2019}), and was further processed with a post facto restoring technique, the multi-object multi-frame blind deconvolution (MOMFBD, \citep{vanoort2005}), which removes for a large part any residual distortions left by AO systems. More details about this data set can be found in \citet{murabito2020}.

\begin{figure}[H]
\,\,\includegraphics[scale=0.4]{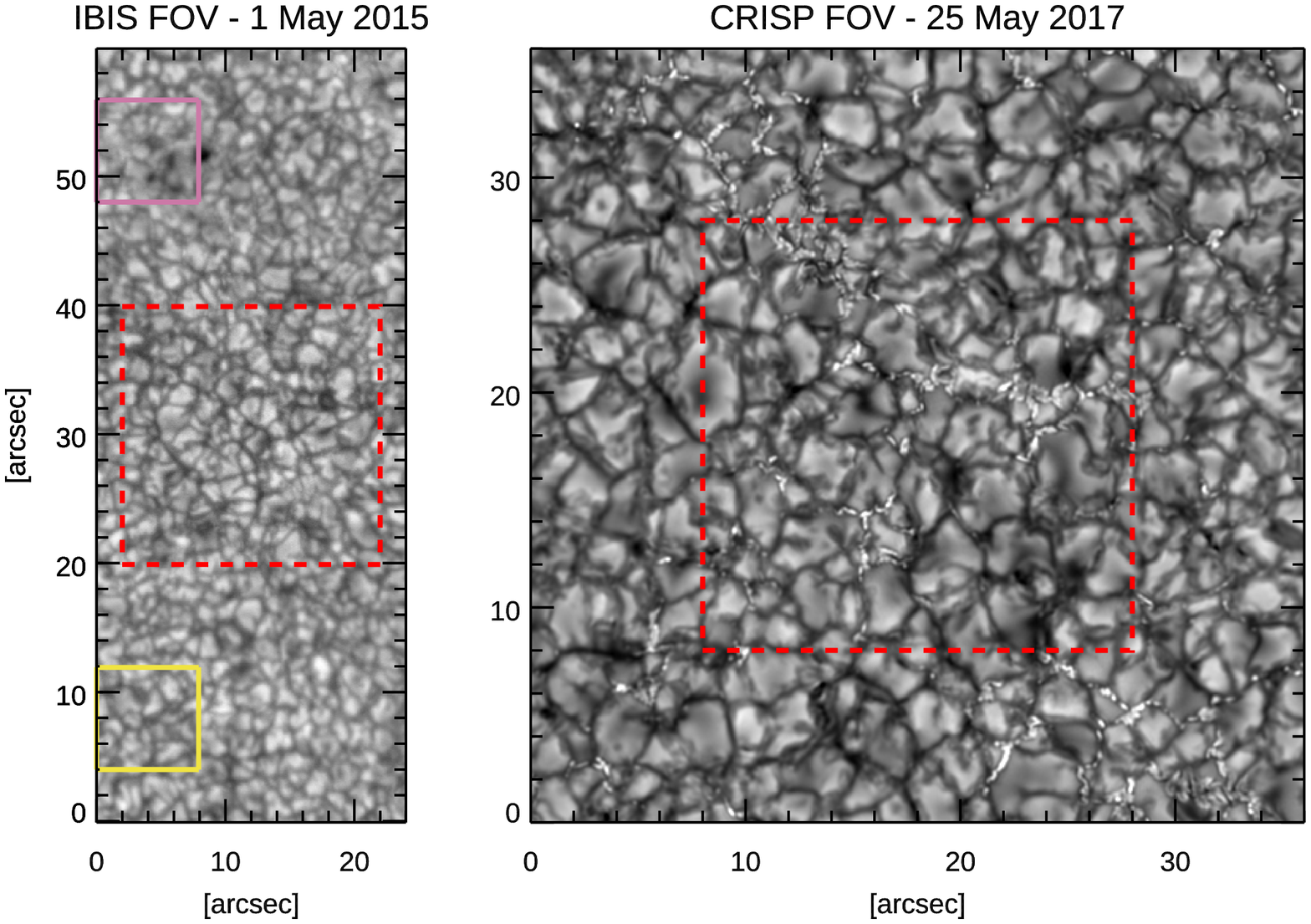}\\
\includegraphics[scale=0.4]{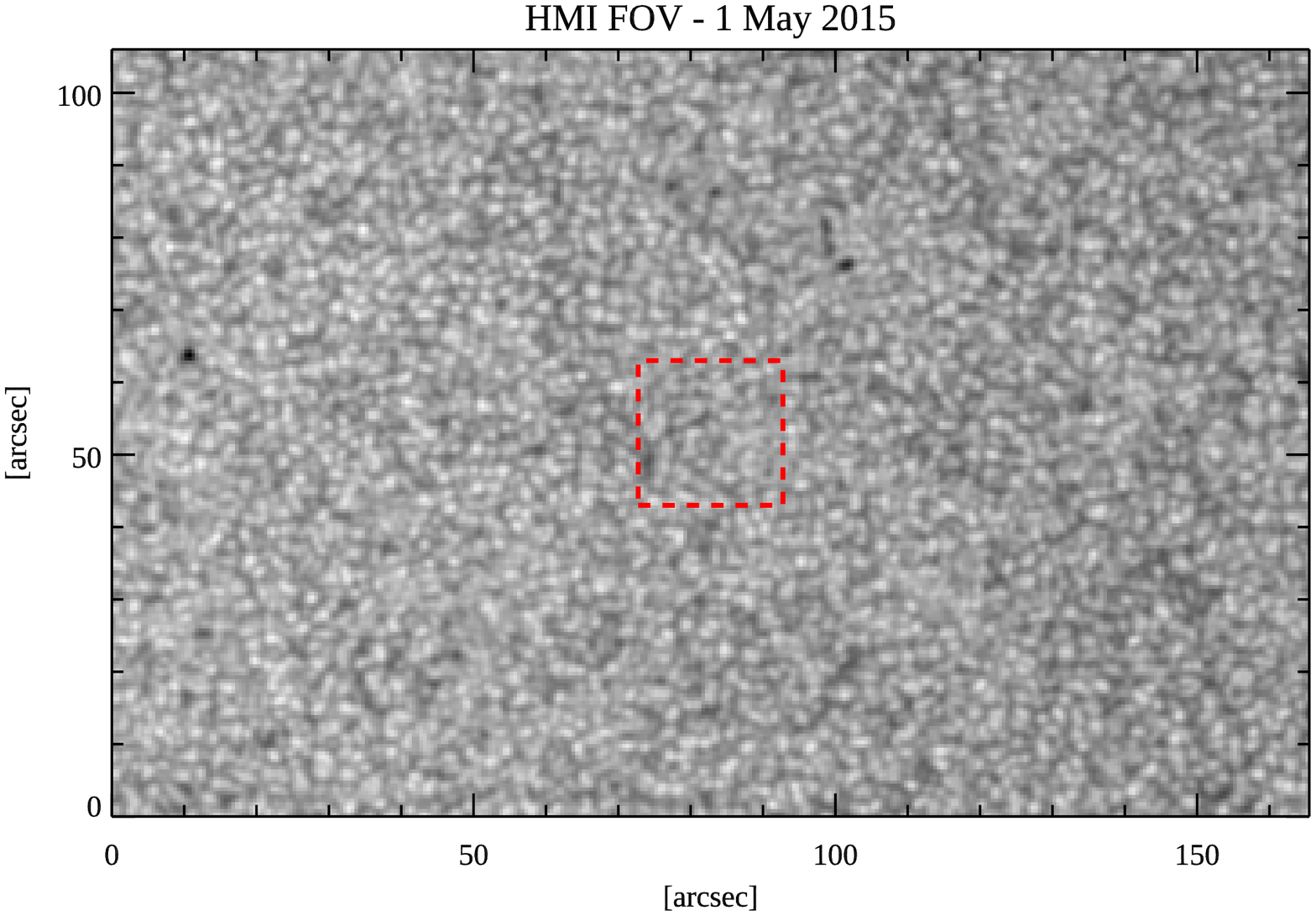}
\caption{Continuum intensity maps of the three data sets used in this work, for IBIS (top-left panel), CRISP (top-right panel), and HMI (bottom panel) observations. The purple and yellow boxes frame the selected IBIS magnetic and quiet sub-FoVs. The dashed red boxes indicate the $20\arcsec \times 20\arcsec$ areas as shown in Figure~\ref{FIG2}. {For the sake of conciseness, we only display the HMI data which are co-temporal with IBIS observations. See Section \ref{sec:datasets} for more details.} \label{FIG1}} 
\end{figure}

Moreover, we analyzed observations acquired on 25 May 2017 with the Crisp Imaging SpectroPolarimeter (CRISP, \citep{scharmer2006}) installed at the Swedish 1 m solar telescope (SST), located in Roque de los Muchachos Observatory, La Palma, Canary Islands (Spain). Similarly to IBIS, CRISP also acquired spectro-polarimetric measurements with a pass-band of $\simeq 4$~pm, performing a scan along the spectral region containing the Fe~I 630.15~nm and 630.25~nm line pair. In this study, we analyzed the Fe~I 630.15~nm data only. The covered FoV was approximately $35\arcsec \times 35\arcsec$. The spatial resolution is about $0.13\arcsec$, corresponding to 90~km on the solar photosphere. The whole data set contains 20 spectral scans. In Figure~\ref{FIG1} (top-right panel), we display the continuum intensity map of the CRISP data set. The observations were processed with the standard reduction pipeline and restored with the dedicated MOMFBD \citep{delacruz2015}. More details about this data set can be found in \citet{Bose2019,Bose2021}.

We complemented our analysis with the study of the full-disk observations taken with the Helioseismic and Magnetic Imager (HMI, \citep{scherrer2012}) instrument on board of the Solar Dynamic Observatory (SDO, \citep{pesnell2012}) satellite {during the IBIS and CRISP observations described above}. In particular, we considered continuum filtergrams, Dopplergrams {and line-width maps derived from the measurements performed along the Fe I 617.3 nm spectral line every 720 s}. We selected a sub-array at the solar disk center, which encompasses an FoV of { $165\arcsec \times 105\arcsec$}, as shown in Figure~\ref{FIG1} (bottom panel) {for HMI data acquired during the IBIS observations}. The spatial resolution is of approximately $1\arcsec$, which corresponds to about 700~km on the photosphere. {In this study, we  only analyzed two individual temporal frames, one relevant to IBIS observations and the other for the CRISP data.}

In order to better visualize the different spatial resolution of each data set considered in our study, we display in Figure~\ref{FIG2} a $20\arcsec \times 20\arcsec$ wide area, cropped at the center of the intensity maps shown in Figure~\ref{FIG1}. Figure~\ref{FIG2} clearly illustrates the increasing spatial resolution from the HMI data set ($1\arcsec$, left panel), to the IBIS data set ($0.16\arcsec$, middle panel), up to the highest resolution in the CRISP data set ($0.13\arcsec$, right panel).

\begin{figure}[H]
	\includegraphics[scale=0.6]{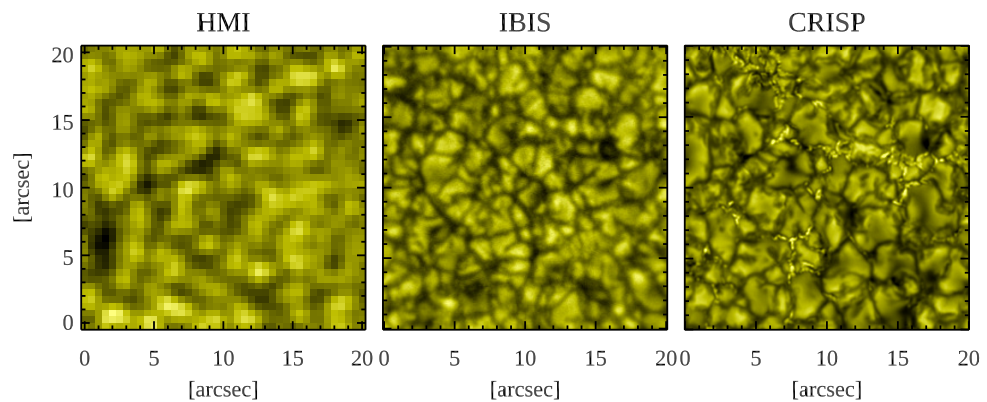}
	\caption{{Zoomed regions relevant to the $20\arcsec \times 20\arcsec$ areas indicated at the center of each panel of Figure~\ref{FIG1}}. 
		\label{FIG2}}
\end{figure}

\subsection{Data Analysis}

Following the approach of \citet{hanslmeier1994}, we computed maps of four physical quantities (hereafter also referred to as parameters) from our observations. These maps show: 

\begin{itemize}
\item Continuum intensity fluctuations, hereafter referred to as $\delta I_{c}$, which are associated with the temperature  differences between hotter granules and cooler intergranular lanes in the observed region;
\item Line-of-sight (LoS) velocity fluctuations, hereafter referred to as $\delta v_{LoS}$, which are related to velocity differences between the upward and downward plasma motions in the studied region, and are encoded in the Doppler shift of the observed spectral lines;
\item Full width at half maximum ($FWHM$) fluctuations, hereafter referred to as $\delta F$, which convey information about changes of the non-thermal motions in the observed region, contributing to the broadening of the spectral line profiles, as for turbulent motions (e.g., \citep{ishikawa2020});
\item Bisector variations of the line profiles associated with the vertical velocity gradients in the observed region; these variations are evaluated using the definition of the line asymmetry parameter $A$ reported in \citet{Hanslmeier1990}, and are hereafter referred to as $\delta A$. 
\end{itemize}

In particular, we computed the maps of $I_c$, $v_{LoS}$, $FWHM$, and $A$ from each observation and estimated the difference (hereafter also referred to as residual) between the values obtained at each point in the map and the mean of values over the whole map. For the HMI set, we could only compute $\delta I_{c}$, $\delta v_{LoS}$, {and} { $\delta F$} from the available data. For the IBIS and CRISP series, we fitted the Stokes~I line measurements at each point in the observation with a linear background and a Gaussian-shaped line, and used the results to deduce the $v_{LoS}$ from the Doppler shift of the centroid of the fitted line, and the $FWHM$ and bisector of the observed line.

We derived the $\lambda $ exponents of the four quantities introduced above, following Equation~2 in \citet{hanslmeier1994}, which is also reported here for reference:\vspace{-6pt}
\begin{equation}
\lambda = \lim_{n \to \infty} \frac{1}{n} \sum_{i=0}^{n-1} \log_{e} |f'(x_{i})|
\label{EQ2}
\end{equation}

\noindent
where $f'(x_{i})$ is the fluctuation of the physical quantity $x$ describing the studied system at the point $i$ of its representation, under the assumption of the ergodic condition mentioned in Section~\ref{sec1}. Here, we remark that this exponent is only a proxy of the real Lyapunov exponent, when the computation procedure is applied to solar images as described in \citet{hanslmeier1994}. This is why we prefer to call it a pseudo-Lyapunov exponent.

In particular, we applied Equation~(\ref{EQ2}) and considered the relation between the values obtained by summing up the residuals of the computed quantities for an increasing number of points in the analyzed map. We then performed a linear fit to the data in the flat asymptotic region of the relation between the values derived from the application of Equation~(\ref{EQ2}) and the number of summed pixels. We assessed the $\lambda$ value by the intercept of the linear fit model applied to the data obtained from each observation and computed the mean of all the $\lambda$ values derived from the observations available for each data set. This mean value is the quantity reported in the tables presented in the following. 

We also analyzed the Stokes~V measurements of the IBIS data set to derive the circular polarization (CP) map shown in Figure~\ref{FIG3}, following the approach of \citet{murabito2017,murabito2019}. We used this map to select regions employed to study the influence of small magnetic fields concentrations on the $\lambda$ exponents. In particular, we selected two sub-FoVs of $8\arcsec \times 8\arcsec$ (corresponding to $100 \times 100$ pixels) near a small magnetic concentration and a dead-calm (hereafter referred to as quiet) area \citep{deadcalm2012}. 
These two sub-FoVs are shown in Figure~\ref{FIG1} (panel a) by the purple and yellow boxes, respectively, and in Figure~\ref{FIG3}. They lie at the same distance from the edges of the IBIS FoV in order to have the same correction effect on the image given by the AO system, and consequently a similar spatial resolution.

\begin{figure}[H]
	\includegraphics[scale=0.6]{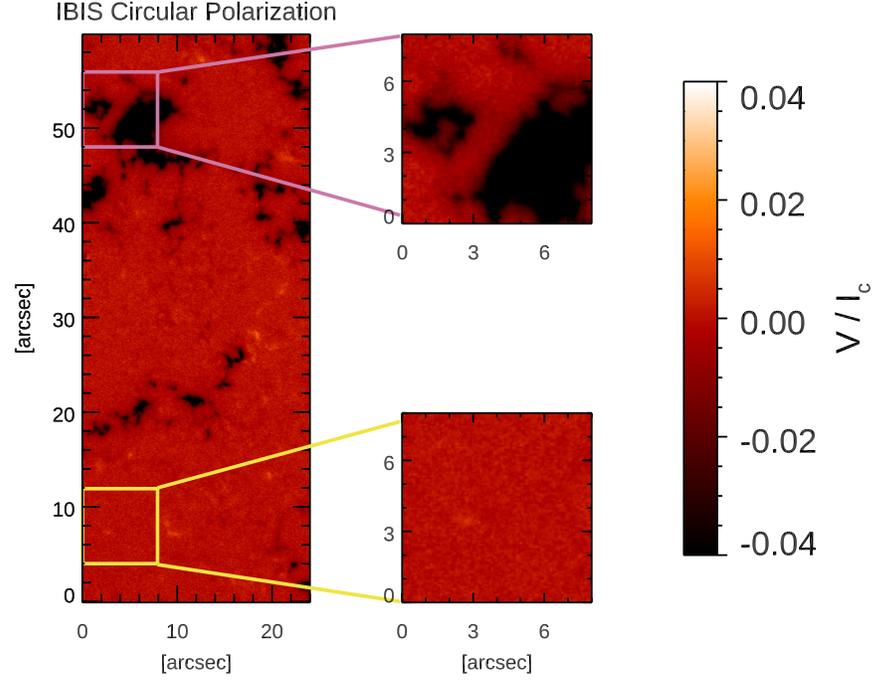}
	\caption{Circular 
polarization (CP) map relevant to the IBIS data set and the magnifications of the magnetic and quiet selected regions.
	\label{FIG3}}
\end{figure}

In order to avoid effects in the estimate of $\lambda$ due to coherent features in the observations, we considered the residuals of the analyzed quantities by performing a random sampling of all the pixels in the FoVs of the computed maps. The large sample of pixels in our observations, which is from 100 to 500 times larger than that considered by \citet{hanslmeier1994}, encodes the properties of a large number of convective cells at different evolutionary stages.  

We estimated the uncertainty of the $\lambda$ exponents derived from IBIS and CRISP observations by computing the standard deviation of the $\lambda$ values obtained from analysis of all the data available for those sets. This uncertainty accounts for the effects of the variable seeing during observations and of the limited resolution of the data. We computed the mean of all the uncertainties, which are reported in the tables. It should be noted that we obtained the values reported in the tables by excluding from the analysis the IBIS scans that were clearly degraded by the seeing or affected by residual degradation over the FoV. For the seeing$-$free HMI observations, we estimated the uncertainty of the computed $\lambda$ by the error of the linear fit model applied to the data.

In contrast to \citet{hanslmeier1994}, we also computed the residuals of the four physical quantities introduced above along a time series of observations, in order to better account in our estimates for the  definition of the Lyapunov exponents $\Lambda$ and their notation in \citet{baker1992}. 
In particular, we computed the residuals of the $I_c$, $v_{LoS}$, $FWHM$, and $A$ at any pixel between the maps derived from observations taken at different times in the IBIS data set. 
We divided the residual values by the time elapsed between the observations represented by the compared maps. We considered observations taken in sequence over a time interval encompassing the typical lifetime of solar granules.

The method we applied is represented by the following equation:
\begin{equation}
 \lambda (\delta t) = \frac{1}{\delta t} \, \lim_{n \to \infty} \frac{1}{n} \sum_{i=0}^{n-1} \log_{e} |\mathcal{M}_{t}-\mathcal{M}_{t_1}|
\label{EQ3}
\end{equation}
where $\mathcal{M}_t$ and $\mathcal{M}_{t_1}$ indicate the maps of the $I_{c}$, $v_{LoS}$, $F$ and $A$ obtained at time $t$ and $t_1$, respectively, $\delta t =t-t_1$, $t_1=t-\tau*j$ if $\tau$ is the cadence (in seconds) of the IBIS data (see Table~\ref{table1}), and $j$ indicates the number of temporal steps between the two compared maps.  
In our estimates, $t$ is the time corresponding to the observation with higher contrast, details of which can be found below, and $j$ goes  from 1 to 10 to account for the typical lifetime of granules (see Section~\ref{sec1}) and the cadence of the analyzed observations. We divided the residuals maps by the time (in seconds) elapsed between the compared  observations, in order to obtain a quantity representative of $\Lambda$ expressed in [s$^{-1}$], as for the definition of the Lyapunov exponents $\Lambda$.

It is worth mentioning that visual inspection suggested that the quality of the IBIS and CRISP observations slightly varies with time.~Therefore, in order to identify the observations with the highest quality in each data set, we computed the image contrast $C$, which is a quantity often employed to evaluate the quality of high-resolution observations. In particular, for each image pixel $i$, we defined the pixel contrast $C_{i}$ as $C_{i} =I_{i}/\bar I$, where $I_{i}$ is the continuum intensity at the image pixel $i$ and $\bar I$ is the mean of the continuum intensity in all the image pixels. Then, we derived the maximum and minimum values of the pixel contrast in each image, $C_{max}$ and $C_{min}$, respectively, and computed for each image the quantity $C=C_{max}-C_{min}$. The largest value of $C$ provides an indication for the best quality scan of each data set. We refer to $C$ in the following.

\section{Results}
\label{sec:results}

Figure~\ref{FIG5new} shows the values for the $\lambda$ exponents  derived from $\delta I_{c}$ fluctuations in the continuum intensity images, {displaying} up to 10,000 image pixels in one frame from each data set. It can be clearly seen that the three exponents converge to negative values. Moreover, for all the computed exponents, there is a flat region that is representative of stable convergence, which is usually reached when using more than about 2000 image pixels for the calculation. We found such a behavior for all the  $\lambda$ exponents computed in this work, so that in the following, we plot the variation of the $\lambda$ exponents computed up to 4000 image pixels only.

\begin{figure}[H]
\includegraphics[scale=0.45]{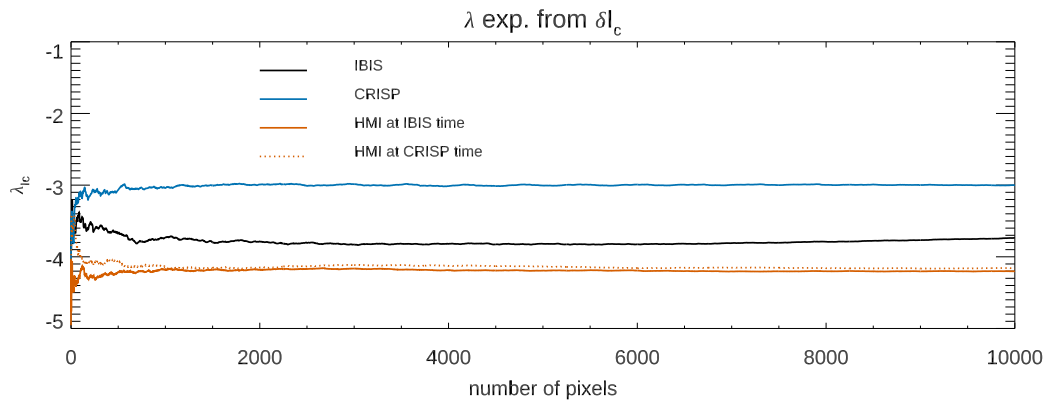}
\caption{{$\lambda$ 
exponents from $\delta I_{c}$ fluctuations derived from one continuum intensity image extracted from each of the three analyzed data sets. See text for more details.}  
\label{FIG5new}}
\end{figure}

The values of the $\lambda$ exponents in Figure~\ref{FIG5new} differ from each other and depend on the analyzed data set. In order to investigate whether the differences among the computed exponents depend on the diverse spatial resolution of the analyzed observations or other data characteristics, e.g., the variable image contrast $C$, we analyzed the relationship between the $\lambda$ exponents computed in each continuum intensity image of the available IBIS and CRISP observations and their contrast $C$. Figure~\ref{FIG5} (panel a and panel b) shows the variation of the 
$\lambda$ exponents derived from $\delta I_{c}$ fluctuations in the continuum intensity images characterized by the highest and lowest $C$ values in the IBIS and CRISP series. The dashed areas indicate the uncertainties associated with our $\delta I_{c}$ estimates in the IBIS and CRISP data sets, which are $\pm$0.1 and $\pm$0.05, respectively. Figure~\ref{FIG5} (panel c and panel d) displays a scatter plot between $C$ and $\lambda$ values derived from the $\delta I_{c}$ fluctuations in the IBIS and CRISP data series. The arrows point to the values from the scans with the highest and the lowest $C$ in each series. It can be clearly seen that there is an almost linear relation between the image quality, defined by its contrast $C$, and the value of the $\lambda$ exponent. In our study, we analyzed all the observations available from the two data sets, except for the seeing degraded IBIS scans. However, in the following, we only show results derived from the analysis of the observations characterized by the highest $C$ value in each series. Table~\ref{TAB2} summarizes the results obtained.  

Figure~\ref{FIG4} (panel a) shows the variation of the $\lambda$ exponents as a function of the number of image pixels computed from $\delta I_{c}$ in the IBIS, CRISP, and HMI data. It is seen that the convergence of the computed $\lambda$ is to negative values. The values in Figure~\ref{FIG4} (panel a) suggest that the computed exponents depend on the spatial resolution of the analyzed observations, obtaining larger $\lambda$ values for the higher spatially resolved CRISP data with respect to the less resolved HMI and IBIS observations. 
As reported in Table~\ref{TAB2}, the average values of the $\lambda$ exponents with their respective uncertainty range between \mbox{$-4.2 \pm 2.0\times10^{-5}$} and $-3.00 \pm 0.05$ obtained from the HMI and CRISP data sets, respectively. The dependence of the $\lambda$ value on the data resolution suggested by the above findings is also supported by the relationship between the exponents obtained from the IBIS and HMI data, which are taken by sampling the same Fe~I 617.3~nm spectral line. It is worth noting that the small errors in Table~\ref{TAB2} derived from HMI data are considerably lower that those for CRISP and IBIS because the HMI observations are unaffected by seeing degradation. 

Figure~\ref{FIG4} (panel b) displays the $\lambda$ exponents evaluated from $\delta v_{LoS}$ fluctuations. Even in this case, the $\lambda$ exponents stably converge to negative values. These increase with the spatial resolution of the data, ranging between { $-2.4 \pm 7.5\times10^{-5}$} and $-1.02\pm 0.06$ for the HMI and CRISP observations, respectively. 

On the other hand, Figure~\ref{FIG4} (panel c) shows the $\lambda$ exponents obtained from $\delta F$ fluctuations. The values of the $\lambda$ exponents are between $-5.18 \pm 0.06$ and { $-6.2 \pm 1.9\times10^{-5}$} for the CRISP and {HMI} data, respectively. Again, the value of the $\lambda$ exponent from the CRISP data is higher than that from IBIS observations{ which, in turn, is higher than the one estimated from the HMI observations}. Figure~\ref{FIG4} (panel d) shows the $\lambda$ exponents evaluated from $\delta A$ fluctuations in the {CRISP and IBIS} data sets. We found a stable convergence of the computed exponents to negative values from both of the analyzed observations. The values of the $\lambda$ exponents are $-5.12 \pm 0.06$ and $-6.0 \pm 0.4$ for the CRISP and IBIS data,~respectively.

\begin{figure}[H]
	\includegraphics[scale=0.435]{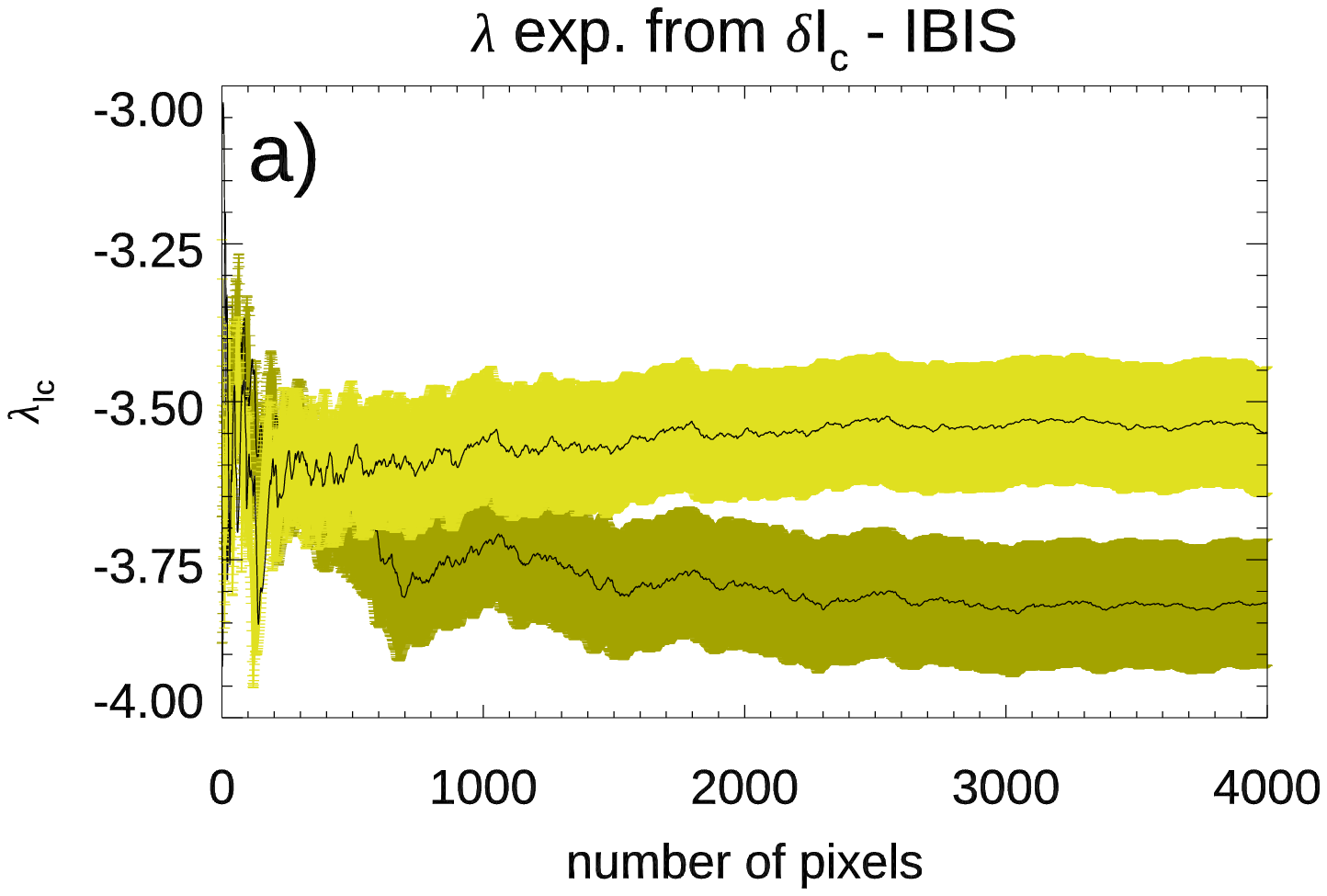}
	\includegraphics[scale=0.435]{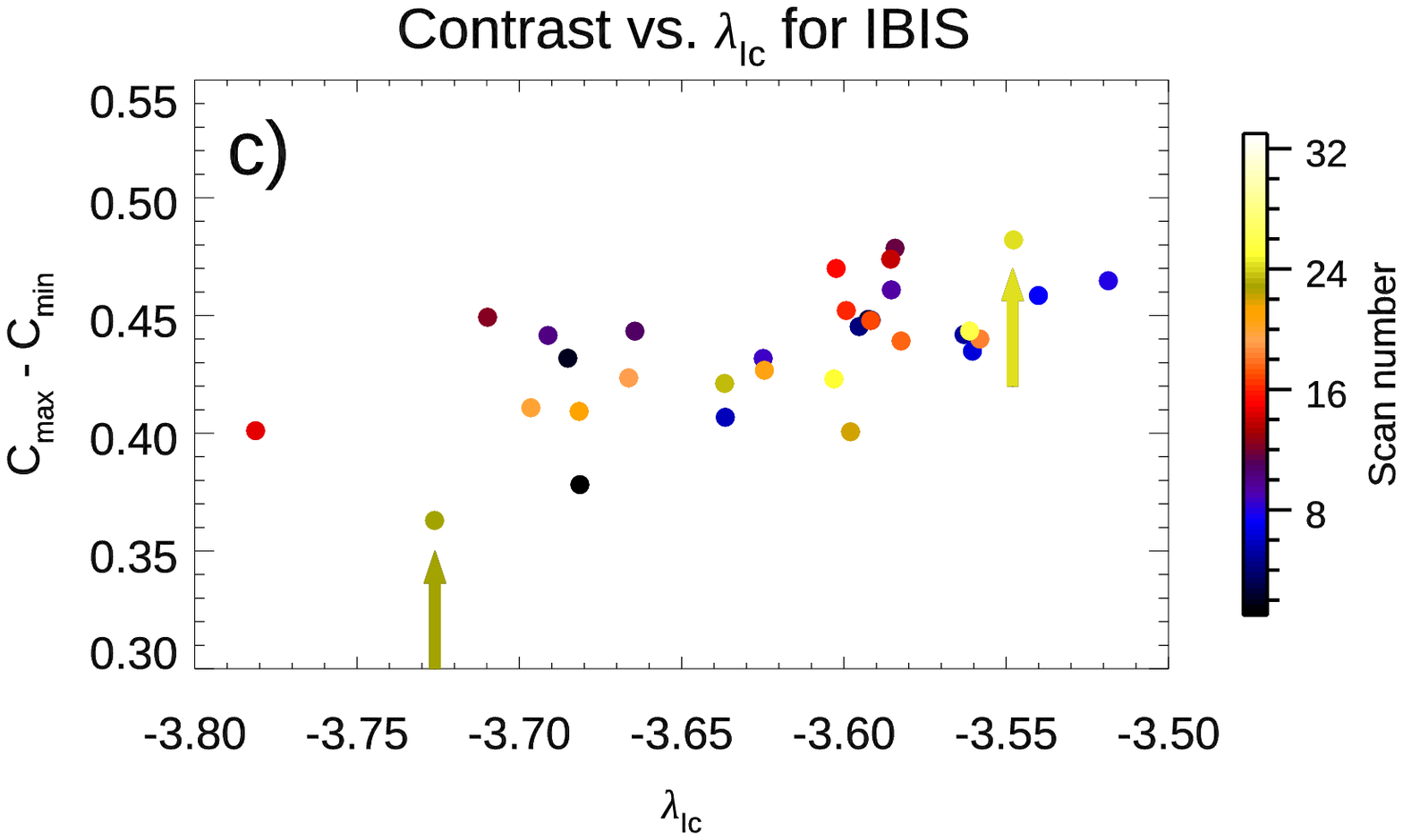}\\%
	\includegraphics[scale=0.435]{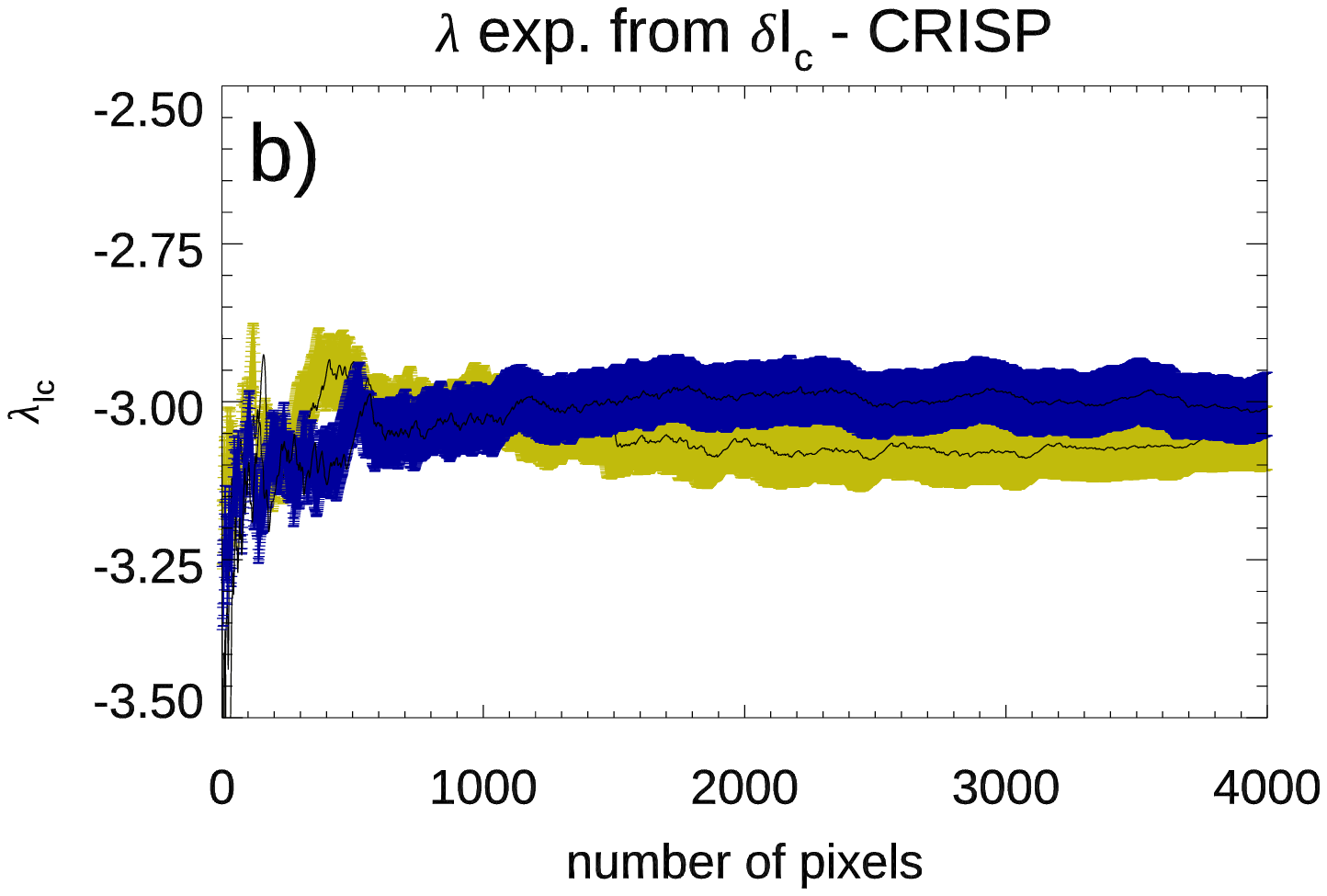}%
	\includegraphics[scale=0.435]{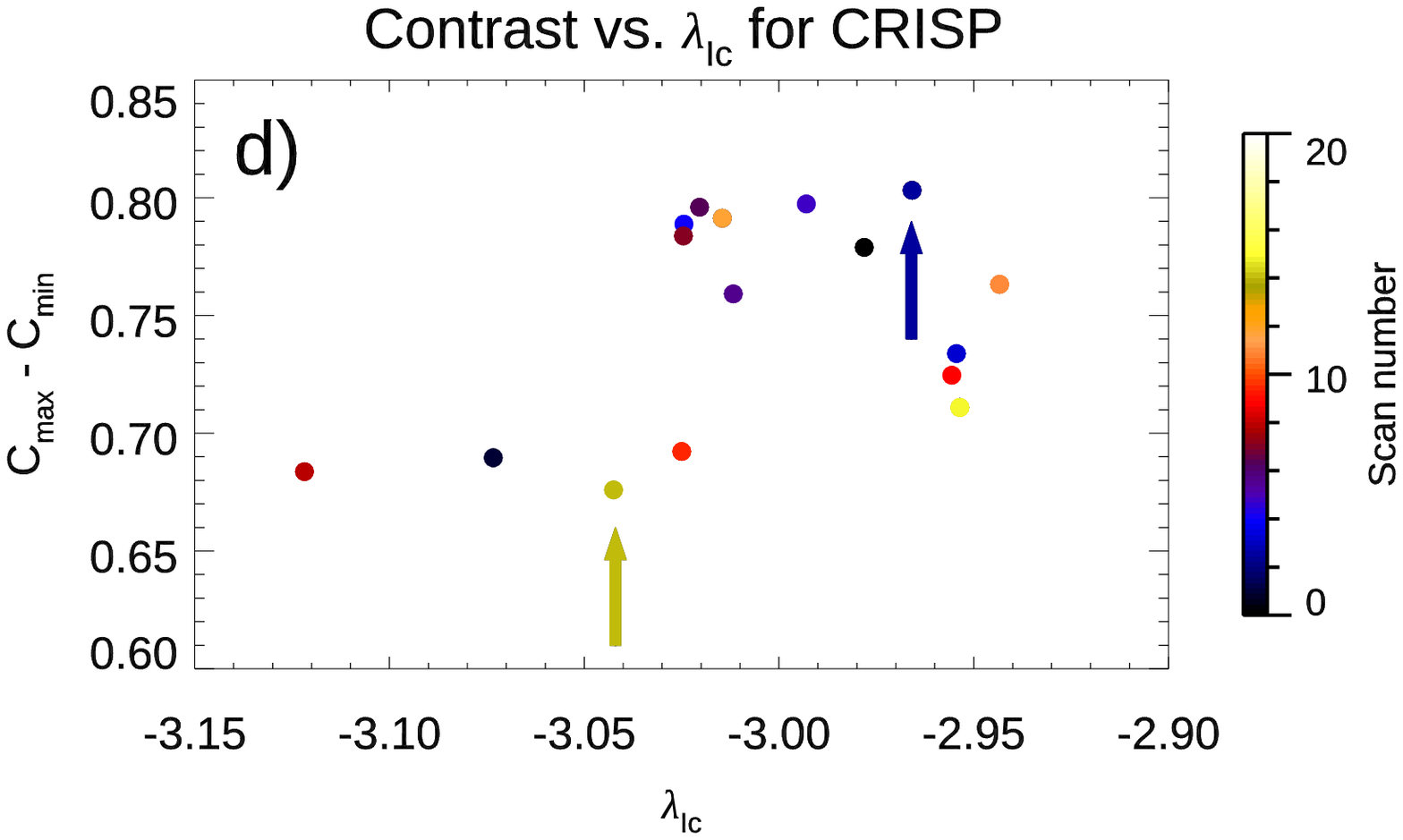}%
	\caption{\textbf{Left}
: $\lambda$ exponents from $\delta I_{c}$ fluctuations evaluated in the highest and lowest $C$ continuum images of the IBIS (panel \textbf{a}) and CRISP (panel \textbf{b}) series. \textbf{Right}: relationship between image contrast $C = C_{max}-C_{min}$ and $\lambda$ exponents from $\delta I_{c}$ fluctuations in the IBIS (panel \textbf{c}) and CRISP \mbox{(panel \textbf{d})~series}. 
	\label{FIG5}}
\end{figure}
\vspace{-6pt}

\begin{specialtable}[H]
\tablesize{\small}
\caption{Average and standard deviation of the $\lambda$ exponents computed from $\delta I_{c}$, $\delta v_{LoS}$,  $\delta F$, and $\delta A$ fluctuations in CRISP, IBIS, and HMI data.}
\label{TAB2}
\setlength{\cellWidtha}{\columnwidth/5-2\tabcolsep-0.4in}
\setlength{\cellWidthb}{\columnwidth/5-2\tabcolsep-0.2in}
\setlength{\cellWidthc}{\columnwidth/5-2\tabcolsep-0.2in}
\setlength{\cellWidthd}{\columnwidth/5-2\tabcolsep+0.4in}
\setlength{\cellWidthe}{\columnwidth/5-2\tabcolsep+0.4in}
\scalebox{1}[1]{\begin{tabularx}{\columnwidth}{>{\PreserveBackslash\centering}m{\cellWidtha}>{\PreserveBackslash\centering}m{\cellWidthb}>{\PreserveBackslash\centering}m{\cellWidthc}>{\PreserveBackslash\centering}m{\cellWidthd}>{\PreserveBackslash\centering}m{\cellWidthe}}
\toprule
        ~              & \textbf{CRISP}               & \textbf{IBIS}           & \textbf{HMI at IBIS Time} & \textbf{HMI at CRISP Time}                     \\ \midrule
        $\delta I_{c}$   & $-$3.00 $\pm$ 0.05  & $-$3.6 $\pm$ 0.1 & $-$4.2 $\pm$ 1.5$\times 10^{-5}$  & $-$4.2 $\pm$ 2.0$\times 10^{-5}$\\ 
        $\delta v_{LoS}$ & $-$1.02 $\pm$ 0.06  & $-$2.1 $\pm$ 0.2 & { $-2.4 \pm 7.5\times 10^{-5}$}  & { $-2.4 \pm 7.0\times 10^{-5}$}\\ 
        $\delta F$  &  $-$5.18 $\pm$ 0.06 & $-$5.6 $\pm$ 0.1 &  { $-6.2 \pm 1.9\times 10^{-5}$}  & { $-6.1 \pm 3.9\times 10^{-5}$} \\ 
        $\delta A$  &  $-$5.12 $\pm$ 0.06 &  $-$6.0 $\pm$ 0.4  &   -  &   - \\ 
       \bottomrule

\end{tabularx}}

\end{specialtable}

Considering the results from the CRISP and IBIS data, it is seen that the relationship between the $\lambda$ values computed from the two data sets is similar for all the investigated quantities, with a clear dependence on the data sets analyzed, and no signatures of a tendency of the computed exponents to positive values. 
Indeed, Figure~\ref{FIG4} reports negative values for all the $\lambda$ exponents derived from our analysis. 
Therefore, all together, the values and trends in Figure~\ref{FIG4} report on fluctuations of the investigated quantities typical of dissipative systems, which tend to point to a less dissipative regime as spatial resolution~increases. 

{It is worth noting that the results presented in Figures \ref{FIG5new} and \ref{FIG4} show that the estimated values increase with the spatial resolution of the analyzed data. However, they also show that the relation among the values obtained from the diverse data depend on other characteristics of the data as well. In particular, we notice that the values estimated from the IBIS observations are closer to the ones derived from HMI data, than those from CRISP observations, in contrast to the diverse spatial resolution between the full-disk HMI and high-resolution IBIS and CRISP data.}

\begin{figure}[H]
	\includegraphics[scale=0.47]{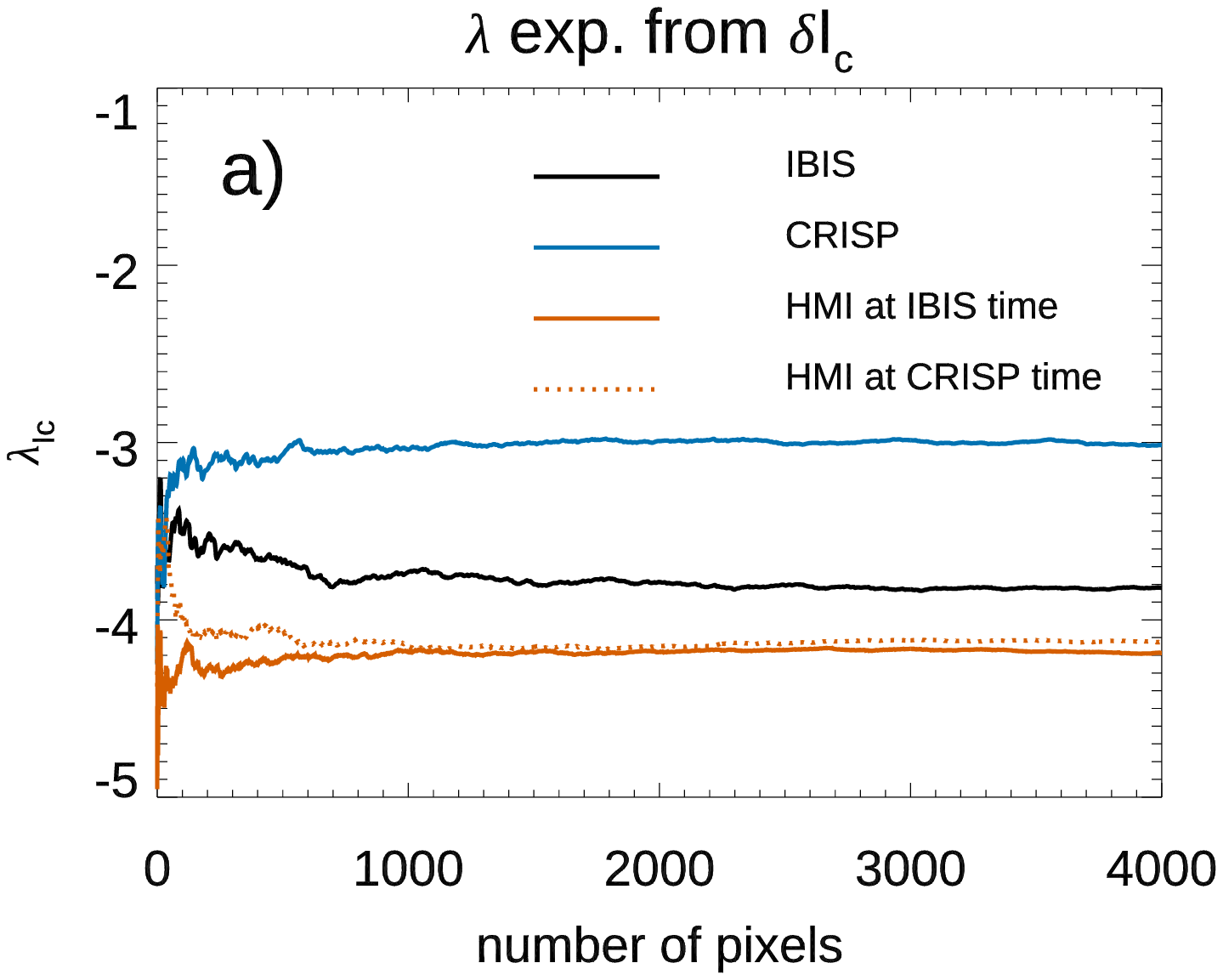}
	\includegraphics[scale=0.47]{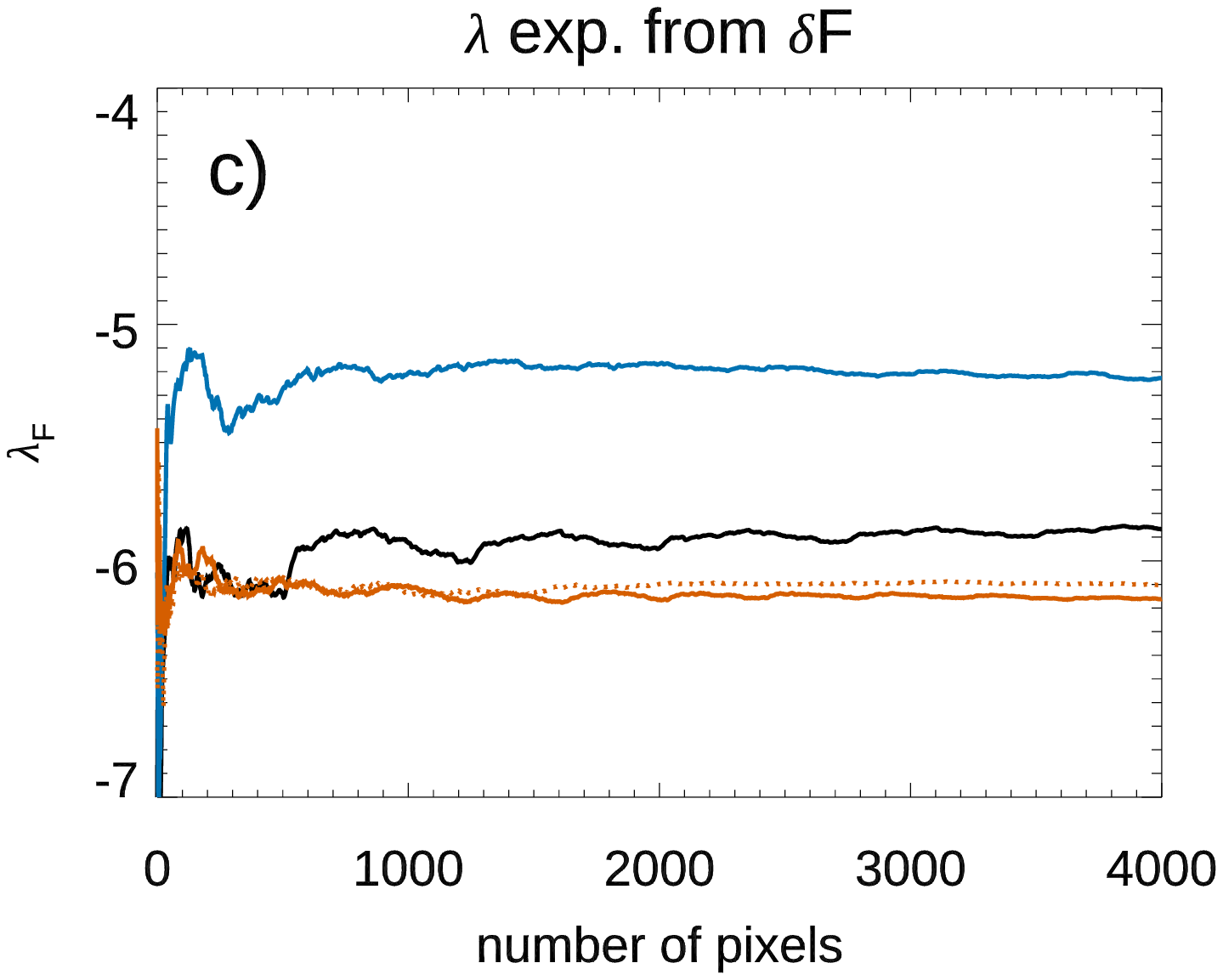}\\
	\includegraphics[scale=0.47]{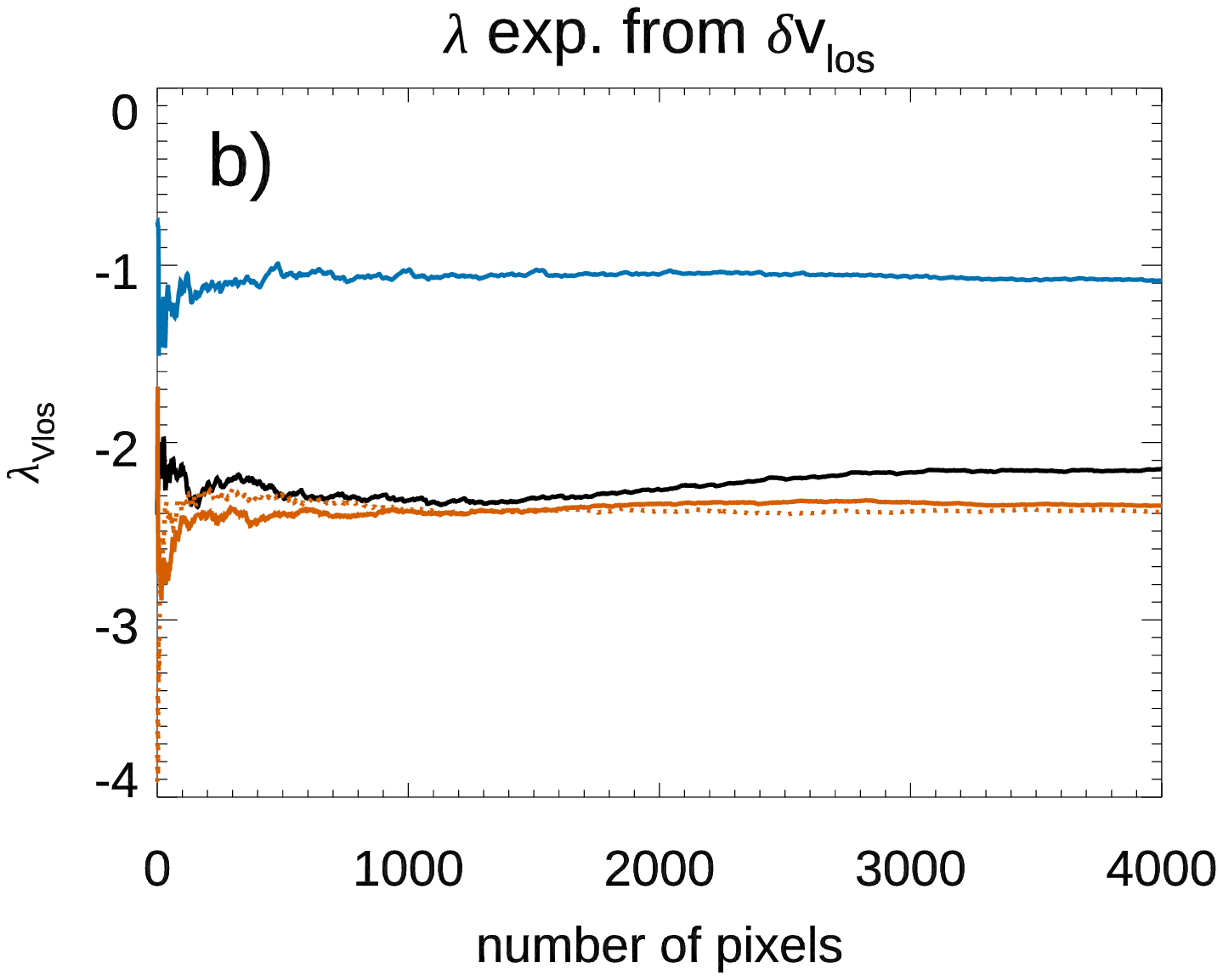}
	\includegraphics[scale=0.47]{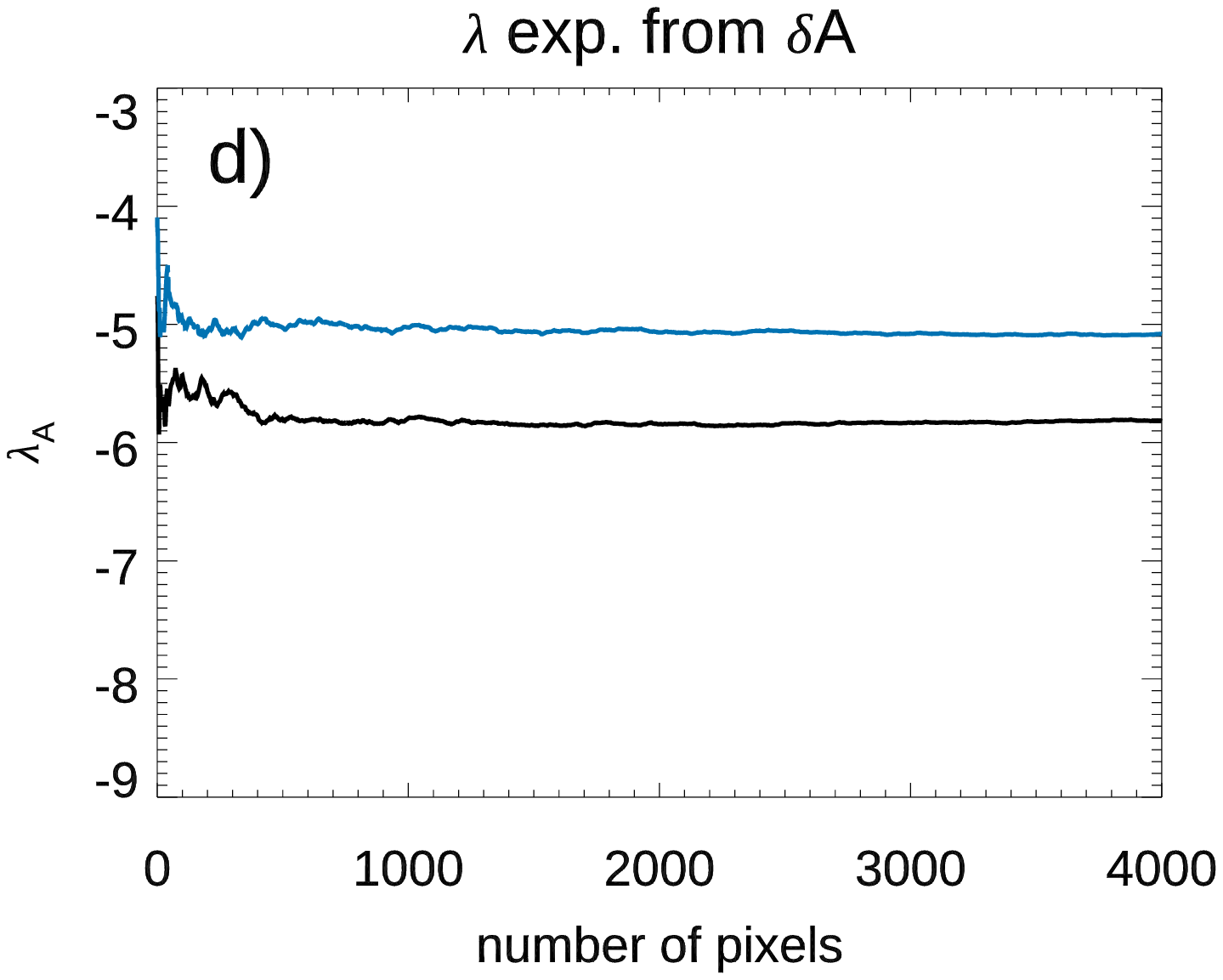}
	\caption{{$\lambda$ 
exponents evaluated from $\delta I_{c}$ (panel \textbf{a}), $\delta v_{LoS}$ (panel \textbf{b}), $\delta F$ (panel \textbf{c}), and $\delta A$ (panel \textbf{d}) fluctuations for IBIS, CRISP, and HMI data sets.}
	\label{FIG4}}
\end{figure}

We note that the IBIS and CRISP data were taken at the Fe~I 617.3~nm and Fe~I 630.15~nm lines, respectively,  which sampled slightly different heights of the photosphere. {Indeed, the formation height of the spectral line sampled by IBIS is slightly deeper in the photosphere than that of the line observed with CRISP. Thus one could expect higher $\lambda$ values in the IBIS data relevant to deeper atmospheric layers.} The height difference could leave signatures, mostly in the $\delta F$ and $\delta A$ derived from these data. {However, it is not the case.} Therefore, the higher value of the $\lambda$ obtained from the CRISP observations with respect to the one computed in the IBIS data could depend on other characteristics of the two observations than their spatial resolution {and the line formation height. However, in} order to investigate if the different spatial resolution is responsible for the gap between the $\lambda$ values derived from IBIS and CRISP data, we degraded the CRISP observations to the expected spatial resolution of the IBIS observation and repeated our analysis.  

Figure~\ref{FIG5bis} displays the variation of the four $\lambda$ exponents obtained from the IBIS and degraded CRISP data, by considering the highest contrast observation in each series. The variation of the $\lambda$ exponents obtained from the full-resolution CRISP best scan were over-plotted as a reference. We found that the degradation of the CRISP data only slightly affects the values of the $\lambda$ exponents obtained from our analysis. In particular, the $\lambda$ exponents computed in the full-resolution (degraded) CRISP observations resulted as $-$3.0 ($-$3.10), $-$1.02 ($-$1.08), $-$5.18 ($-$5.24), $-$5.12 ($-$5.37) from the $\delta I_c$, $\delta v_{LoS}$, $\delta F$, and $\delta A$, respectively. We noticed that only the variation of the exponents derived from $\delta A$ fluctuations exceeded the uncertainty of our measurements. Therefore, we thus concluded that the difference between the $\lambda$ exponents derived from the IBIS and CRISP data cannot be ascribed to the diverse spatial resolution of the data alone, but should be originated by other characteristics of the data.

\begin{figure}[H]
	\includegraphics[scale=0.48]{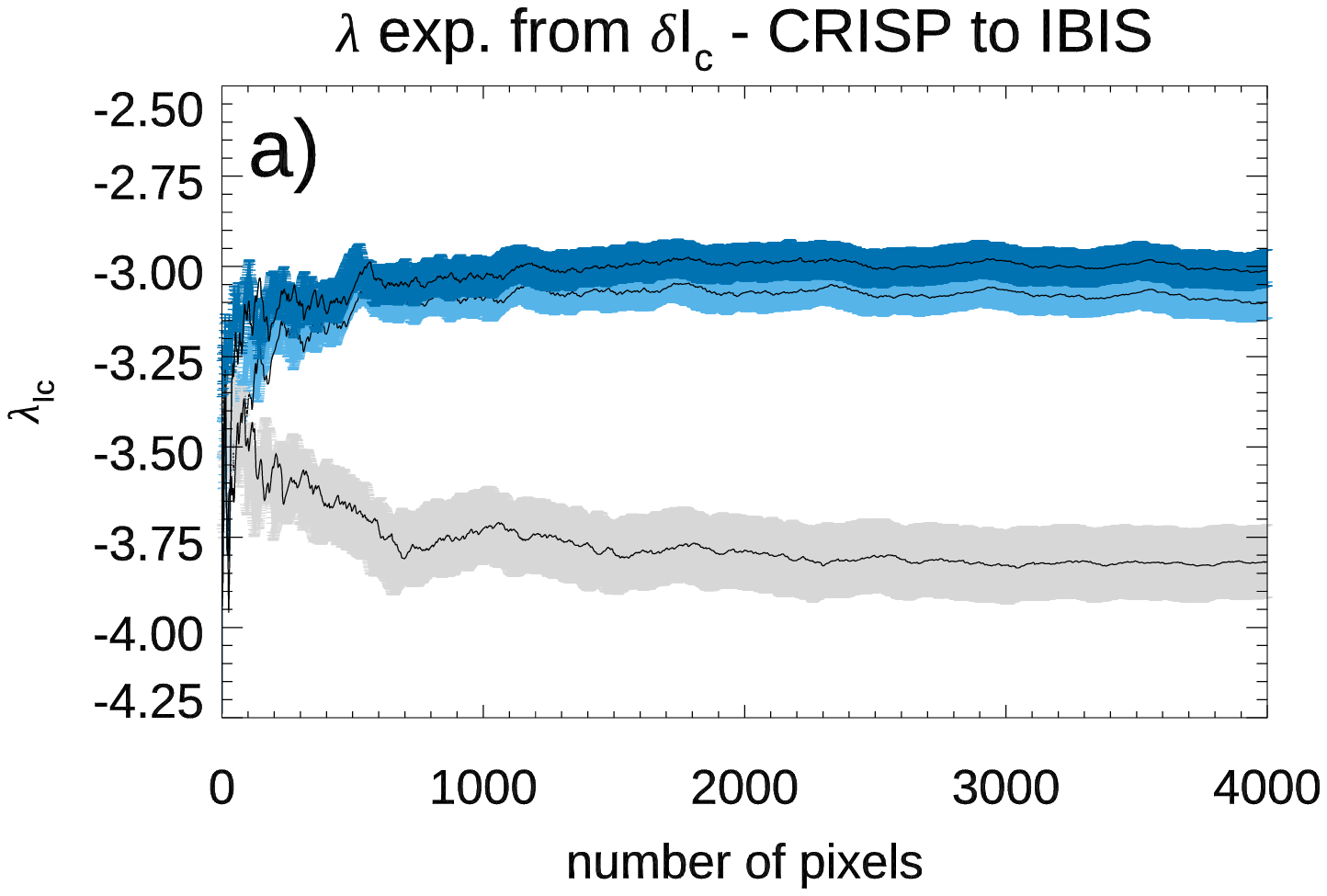}%
	\includegraphics[scale=0.48]{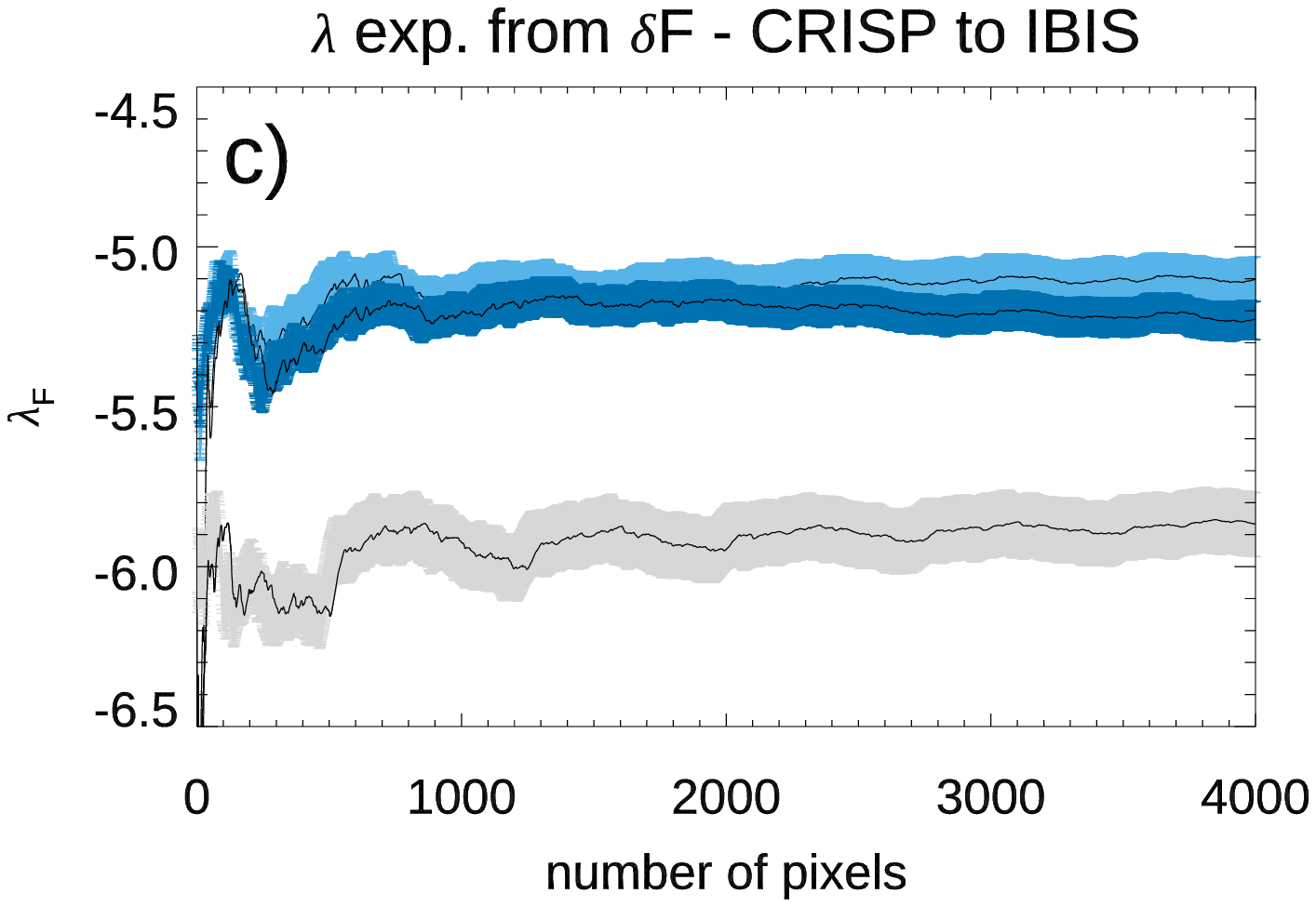}
	\includegraphics[scale=0.48]{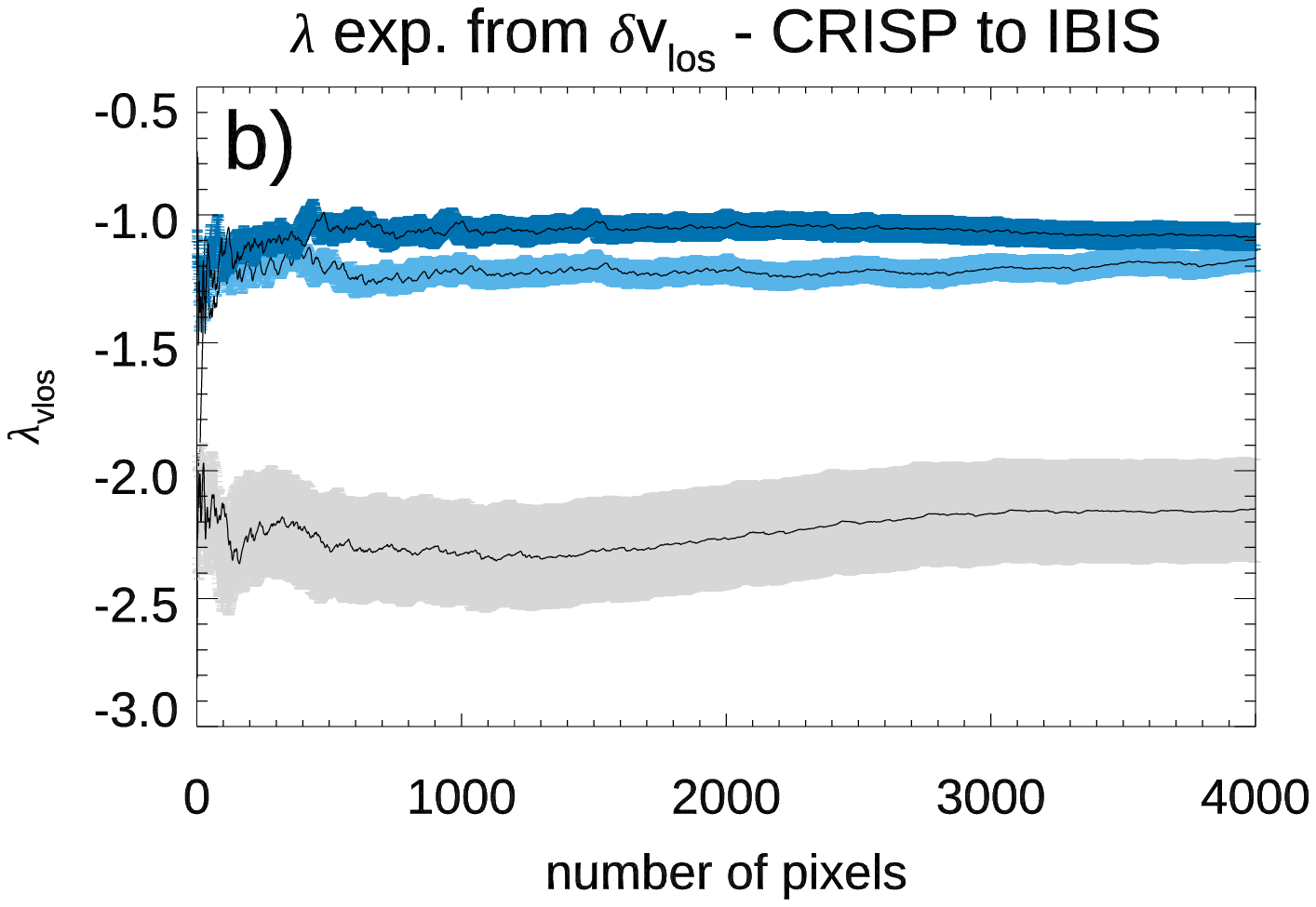}%
	\includegraphics[scale=0.48]{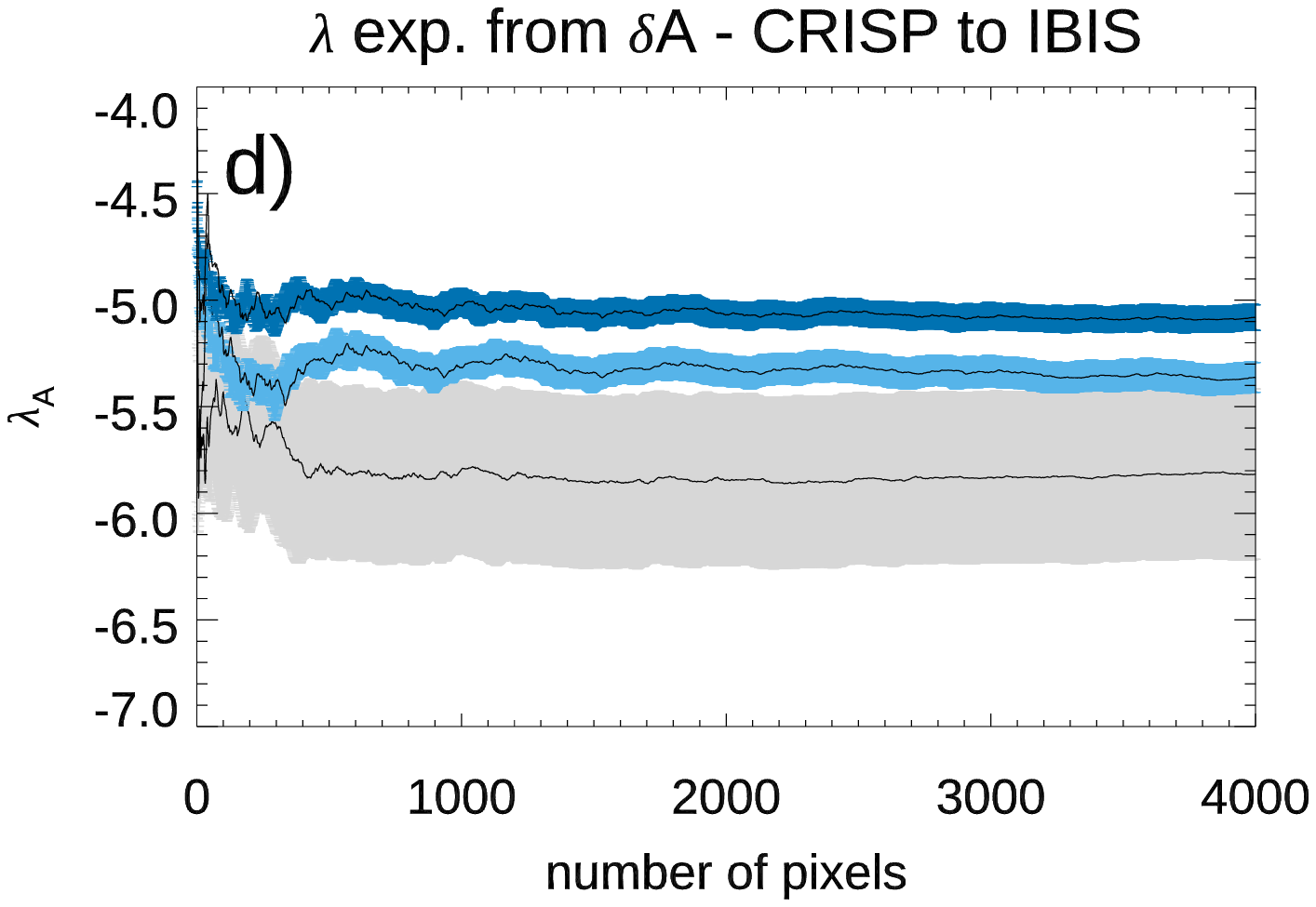}%
	\caption{$\lambda$ 
exponents from $\delta I_{c}$ (panel \textbf{a}), $\delta v_{LoS}$ (panel \textbf{b}), $\delta F$ (panel \textbf{c}), and $\delta A$ (panel \textbf{d}) fluctuations, by considering the highest contrast scans in the IBIS (gray shaded area) and CRISP (blue shaded area) series, and the CRISP degraded data (light blue shaded area). See text for more details. 
	\label{FIG5bis}}
\end{figure}

{In this respect, it is worth noting that the IBIS and CRISP observations differ for the solar features imaged in the FoV. Indeed, the IBIS FoV does not show visible bright points, while the CRISP FoV contains several ``chains'' of these features along the intergranular lanes. These can be clearly noticed in Figure \ref{FIG1} and in more detail, in Figure \ref{FIG2}, thanks to the higher spatial resolution of the CRISP instrument. These features could leave signatures in the various estimated $\lambda$ exponents. In order to investigate if bright points affect the estimated values, we considered the $\lambda$ exponents derived from sub-FoVs including and lacking such solar features. We found that the values derived from the two data samples differ for quantities that only slightly exceed the uncertainty of our estimates reported in Table \ref{TAB2}, but for the values obtained from $\delta F$ fluctuations, which differ for a factor of 10 in terms of the computed uncertainty. The higher $\lambda$ exponent obtained for the $\delta F$ fluctuations in the FoV containing the bright points could be attributed to the turbulent motions \citep{romano2012} reported in these features which induce broadening of the spectral profiles. Contrary to what might have been expected by considering the significantly higher intensity of the bright points with respect to the surrounding solar atmosphere, the values derived from $\delta I_{c}$ fluctuations estimated in regions including chains of bright points only slightly differed from the ones obtained in solar surface areas clearly lacking such features. This could also be the case for the $\delta v_{LoS}$ fluctuations, since downflow motions are usually observed in these features.}

Thus, we suggested that the difference existing between values of the $\lambda$ exponents derived from IBIS and degraded CRISP data may be attributed to the characteristics of the measurements and data processing, such as the level of stray-light in the observations and photometric accuracy of the post facto processing, respectively.

{It is worth noting that the results presented above were derived from the analysis of the whole FoVs and sub-FoVs displayed in Figure \ref{FIG1}. However, we also computed the $\lambda$ exponents on several sub-FoVs with a size comparable to that shown in Figure \ref{FIG1}, which were extracted at different positions of the IBIS FoV. This was to test the accuracy of our results upon the analysis of different solar surface regions. We found that the values of the $\lambda$ exponents estimated from the diverse sub-FoVs all lay within two times the uncertainty of the exponents derived from the analysis of the whole IBIS FoV reported in Table \ref{TAB2}, and within the uncertainty of our estimates for the sub-FoVs in Figure \ref{FIG1} that are listed in \mbox{Table \ref{TAB3}} presented in the next Section.}


\subsection{Further Analyses}

In order to further investigate the dependence of the computed $\lambda$ exponents on the spatial resolution of the analyzed data, we  performed a spatial degradation of the IBIS observations and repeated our analysis on data affected by different degradation levels. To degrade the observation, we used a smoothing function based on Gaussian kernels. We  degraded the maps of the continuum intensity $I_c$, LoS velocity $v_{LoS}$, line $FWHM$ and $A$ derived from the IBIS data to the same spatial scales as used in \citet{hanslmeier1994}, specifically at the following:

\begin{itemize}
\item Subgranular, corresponding to a spatial resolution of 0.36\arcsec{} ($2\times2$ pixels kernel);
\item Granular, corresponding to a spatial resolution of 0.90\arcsec{} ($5\times5$ pixels kernel);
\item Mesogranular, corresponding to a spatial resolution of 5.04\arcsec{} ($28\times28$ pixels kernel).
\end{itemize}

Table~\ref{TAB3} summarizes the results obtained from the analysis of all the IBIS degraded data. 
Figure~\ref{FIG6} (panel a) shows the $\lambda$ exponents computed from $\delta I_{c}$ fluctuations in IBIS degraded data. For comparison, we also plotted the values of the $\lambda$ exponents from the full-resolution IBIS data set and from the HMI observations. 
The $\lambda$ exponents derived from $\delta I_c$ fluctuations are $-3.8 \pm 0.1$, $-4.3 \pm 0.1$, and $-5.6 \pm 0.3$ when degrading the IBIS data with kernels at the subgranular, granular, and mesogranular spatial scales, respectively. There is thus a clear tendency of the $\lambda$ values to increase with the lesser degraded data. 
Figure~\ref{FIG6} (panels b, c, d) display the $\lambda$ exponents derived from the $\delta v_{LoS}$, $\delta F$, and $\delta A$ fluctuations in the IBIS data degraded with the various kernels. We note that the data degradation mostly affects the value of the $\lambda$ exponents computed from $\delta v_{LoS}$ and $\delta A$ fluctuations, i.e., from the quantities most representative of the convective motions of solar plasma at small scales. For the sake of completeness, it is worth mentioning that we obtained similar results using a Fourier filtering of the data.

\begin{specialtable}[H]
\tablesize{\small}
\caption{Average values and standard deviation of the $\lambda$ exponents computed from 
$\delta I_{c}$, $\delta v_{LoS}$, $\delta F$, and $\delta A$ fluctuations in the IBIS data degraded with various Gaussian kernels and in the two sub-FoVs representative of magnetic and quiet regions. See Sect.~3.1 for more details. } 
\label{TAB3}
\setlength{\cellWidtha}{\columnwidth/5-2\tabcolsep+1in}
\setlength{\cellWidthb}{\columnwidth/5-2\tabcolsep-0.25in}
\setlength{\cellWidthc}{\columnwidth/5-2\tabcolsep-0.25in}
\setlength{\cellWidthd}{\columnwidth/5-2\tabcolsep-0.25in}
\setlength{\cellWidthe}{\columnwidth/5-2\tabcolsep-0.25in}
\scalebox{1}[1]{\begin{tabularx}{\columnwidth}{>{\PreserveBackslash\centering}m{\cellWidtha}>{\PreserveBackslash\centering}m{\cellWidthb}>{\PreserveBackslash\centering}m{\cellWidthc}>{\PreserveBackslash\centering}m{\cellWidthd}>{\PreserveBackslash\centering}m{\cellWidthe}}
\toprule
        ~              & \boldmath$\delta I_{c}$               & \boldmath$\delta v_{LoS}$              & \boldmath$\delta F$   & \boldmath$\delta A$               \\ \midrule

        Kernel 2 $\times$ 2 (subgranular)  & $-$3.8 $\pm$ 0.1 & $-$2.2 $\pm$ 0.2 & $-$5.7 $\pm$ 0.1 & $-$6.2 $\pm$ 0.4 \\ 
        Kernel 5 $\times$ 5 (granular) & $-$4.3 $\pm$ 0.1 & $-$2.5 $\pm$ 0.2    & $-$5.8 $\pm$ 0.1 & $-$6.5 $\pm$ 0.4 \\ 
        Kernel 28 $\times$ 28 (mesogranular) & $-$5.6 $\pm$ 0.3 & $-$3.2 $\pm$ 0.3   & $-$6.2 $\pm$ 0.2 & $-$7.1 $\pm$ 0.5 \\ 
        \midrule

        Magnetic sub-FoV & $-$3.9 $\pm$ 0.2 & $-$2.1 $\pm$ 0.3  & $-$5.8 $\pm$ 0.2 & $-$5.9 $\pm$ 0.4 \\ 
        Quiet sub-FoV      & $-$3.6 $\pm$ 0.1 & $-$2.2 $\pm$ 0.3  & $-$5.8 $\pm$ 0.2 & $-$6.2 $\pm$ 0.2 \\ 
\bottomrule
\end{tabularx}}

\end{specialtable}

We noticed that the value of the $\lambda$ exponent derived from the $\delta I_{c}$ fluctuations in the HMI observations lie between those evaluated from the IBIS data degraded at the subgranular and granular scales. This finding suggests that the actual resolution in the degraded IBIS data could actually be slightly worse than the one expected from the applied kernel. On the other hand, the $\lambda$ value from the $\delta v_{LoS}$ fluctuations in HMI data is between the values of $\lambda$ exponent computed in the IBIS data degraded with the granular and mesogranular kernels, as expected based on a dependence of the computed value on the spatial resolution of the data. {As expected, the significantly lower resolution of the HMI data does not allow highlighting any contribution from the sub-granular structures. Indeed, the HMI maps only show granules, without any further spatial detail.}

\begin{figure}[H]
	\includegraphics[scale=0.4]{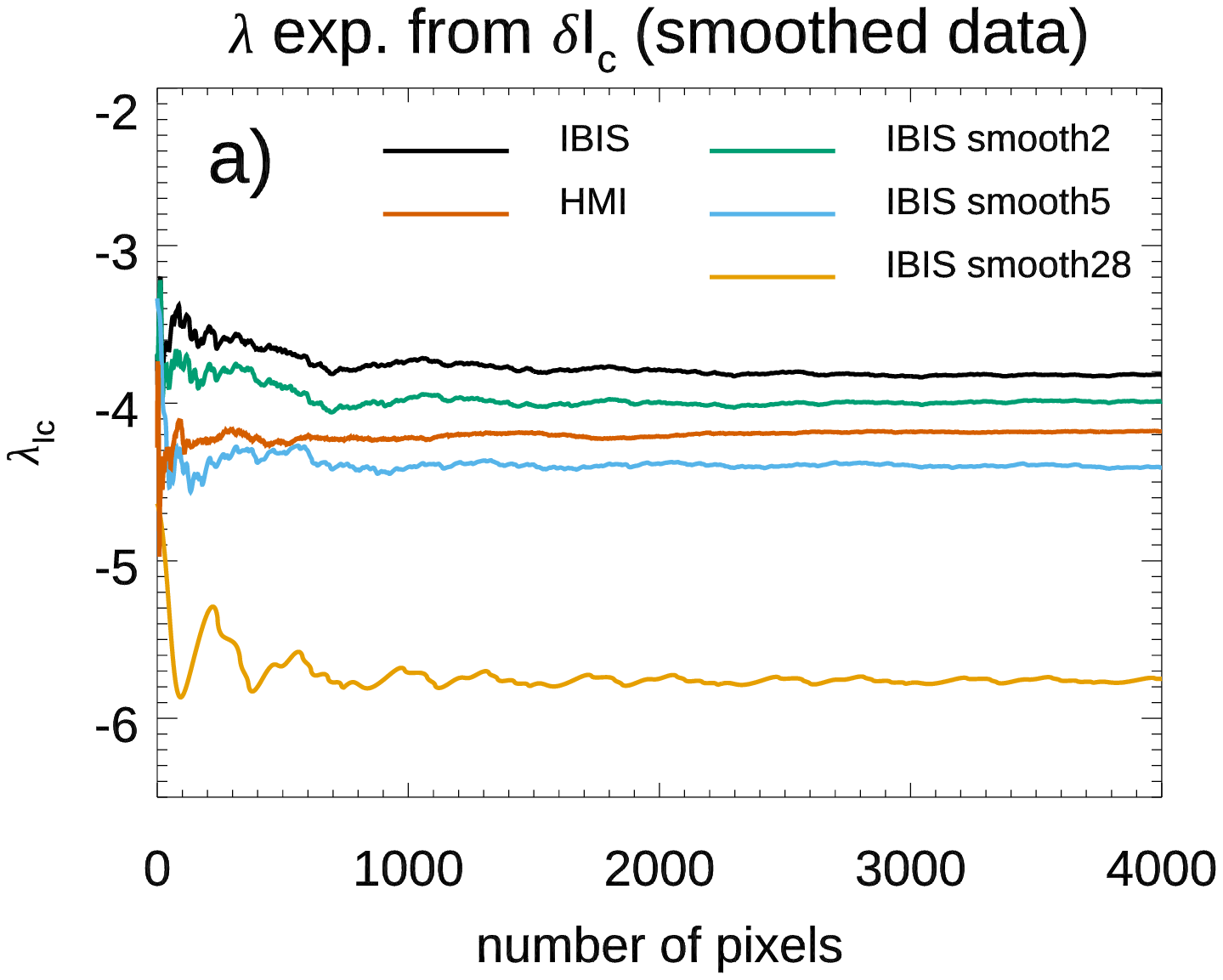}%
	\includegraphics[scale=0.4]{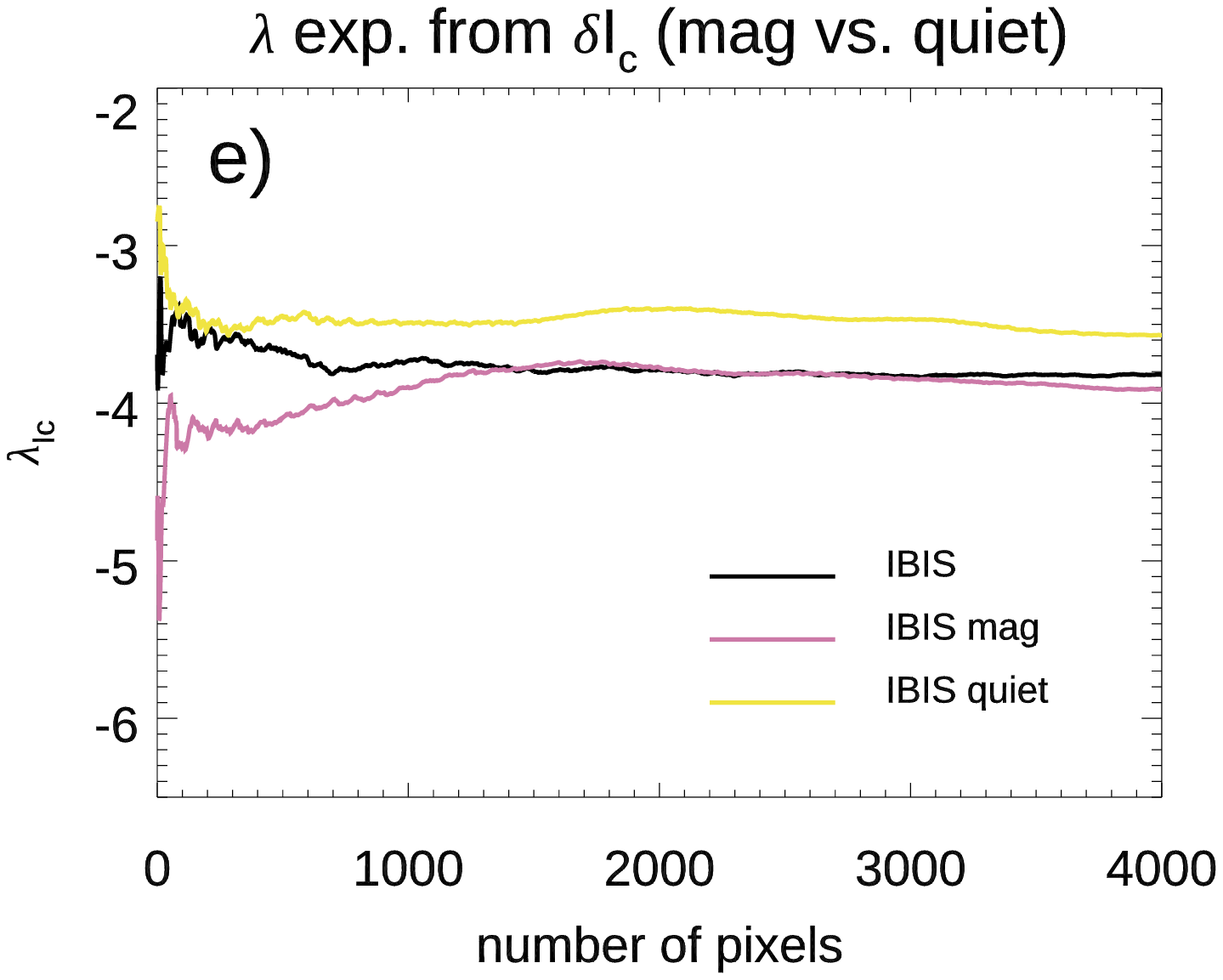}\\
	\includegraphics[scale=0.4]{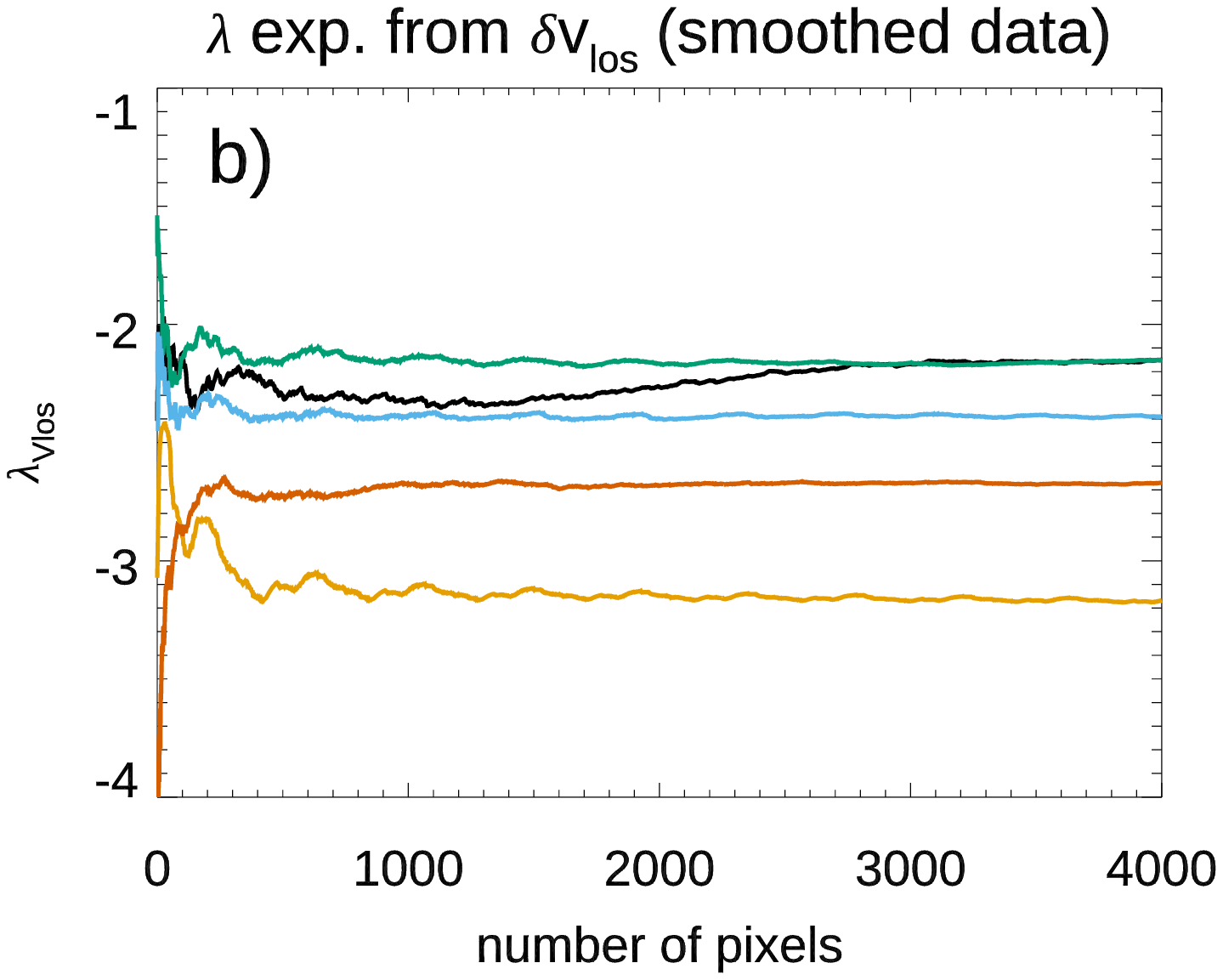}%
	\includegraphics[scale=0.4]{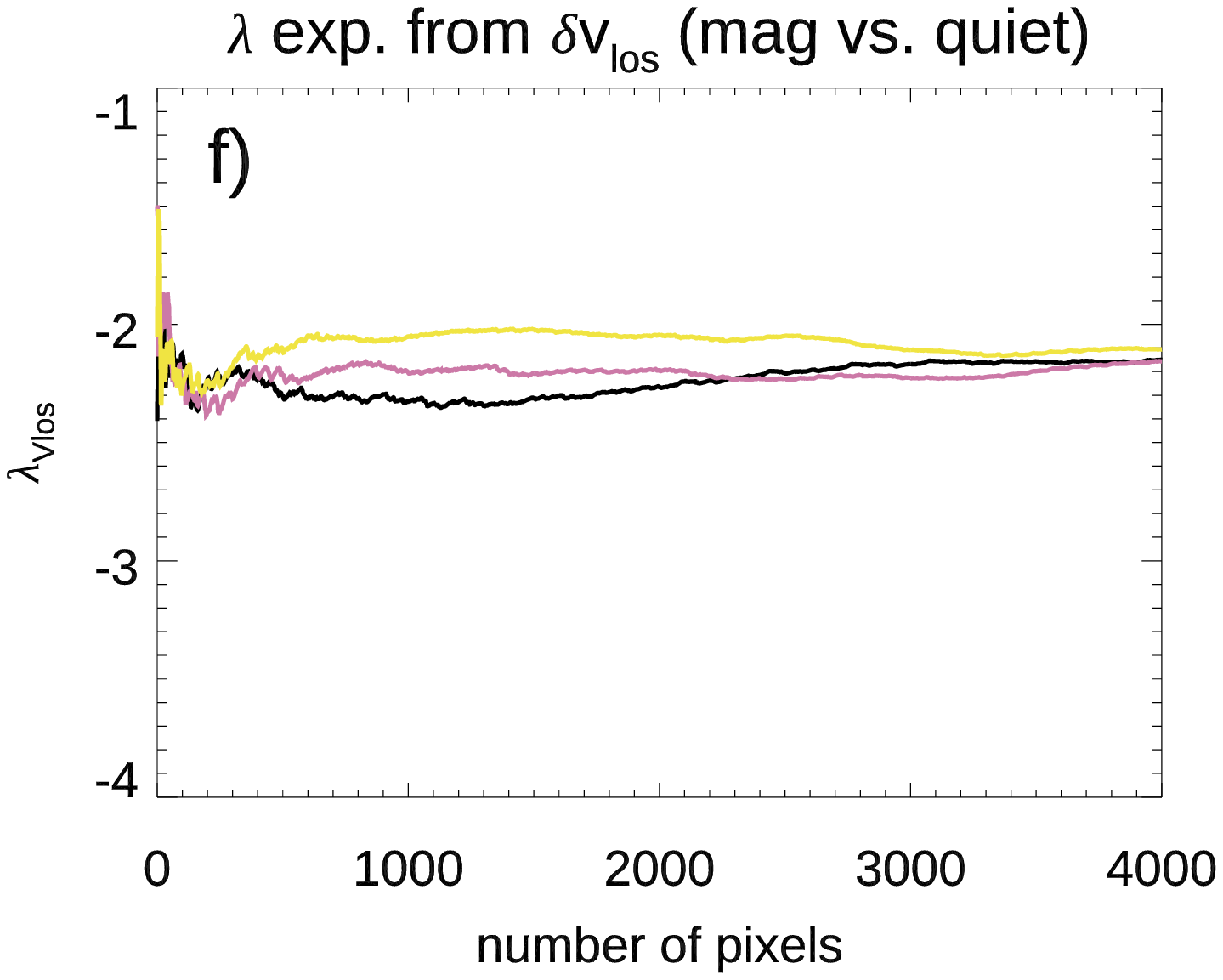}
	\includegraphics[scale=0.4]{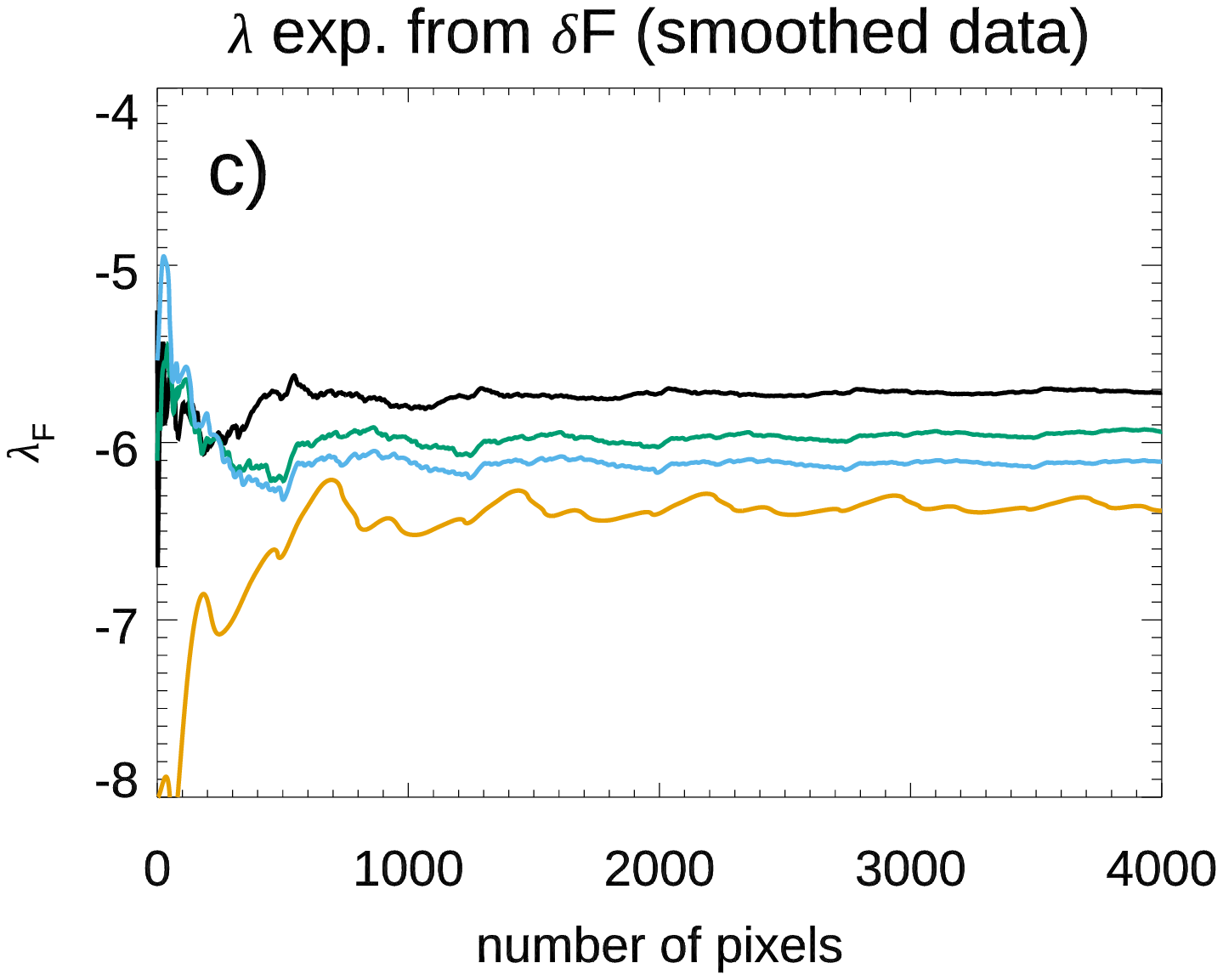}%
	\includegraphics[scale=0.4]{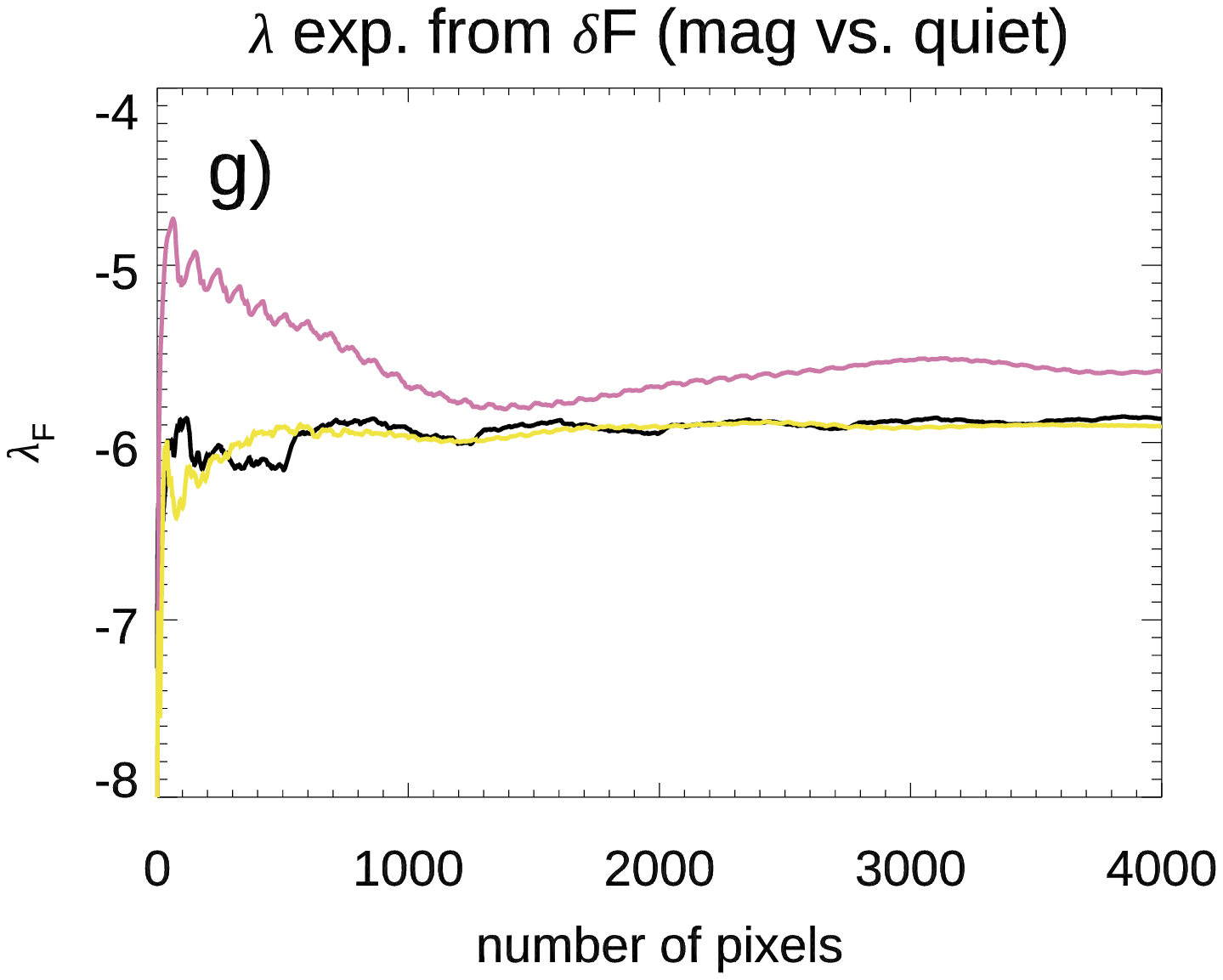}
	\includegraphics[scale=0.4]{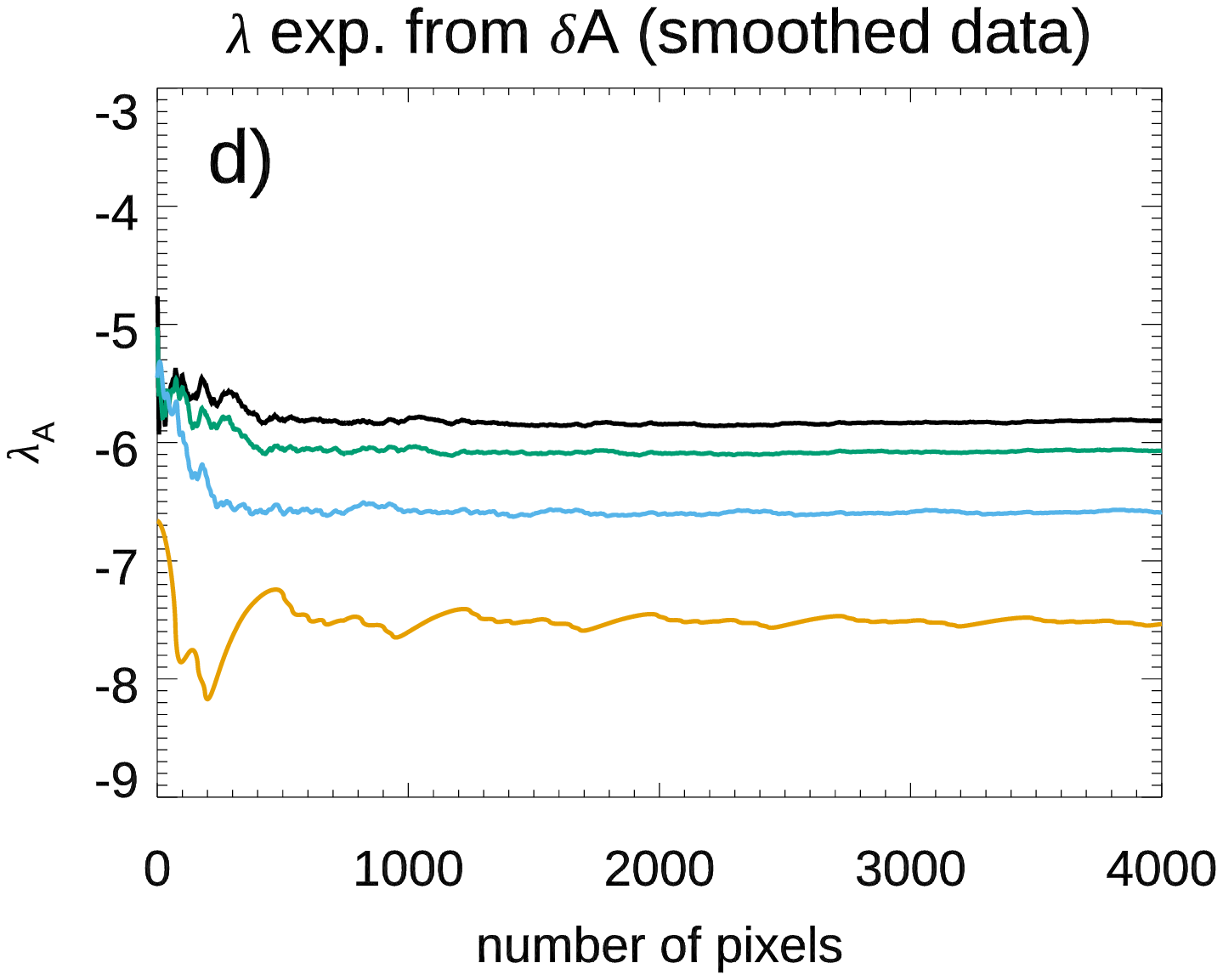}%
	\includegraphics[scale=0.4]{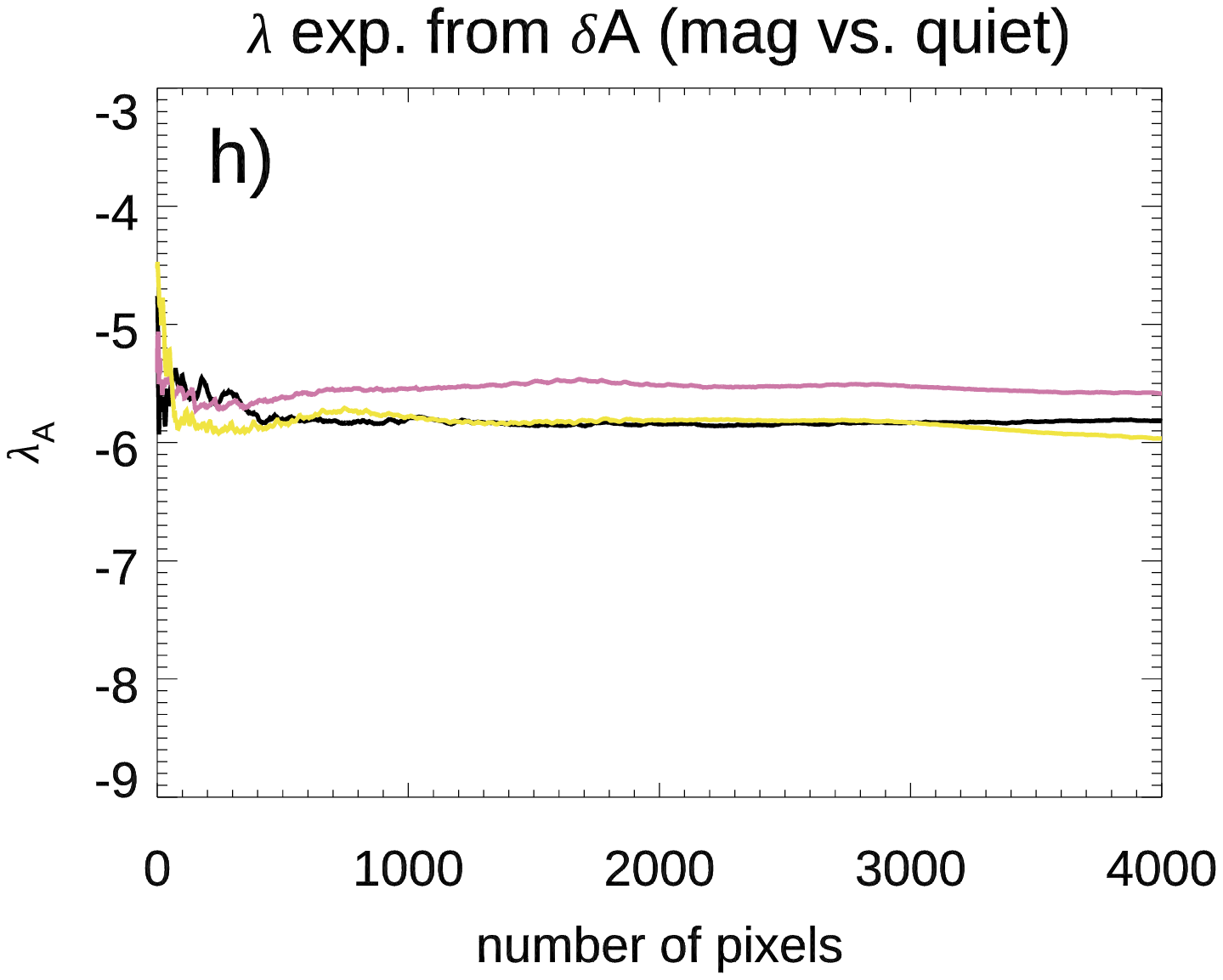}
	\caption{\textbf{Left}
: $\lambda$ exponents computed  from $\delta I_{c}$ ({\bf a}),  $\delta v_{LoS}$ ({\bf b}), $\delta F$ ({\bf c}), and $\delta A$ ({\bf d}) fluctuations in the IBIS data smoothed to subgranular ($2\times2$ pixel smooth), granular ($5\times5$ pixel smooth), and mesogranular ($28\times28$ pixel smooth) spatial scales. \textbf{Right}: $\lambda$ exponents evaluated from $\delta I_{c}$ ({\bf e}), $\delta v_{LoS}$ ({\bf f}), $\delta F$ ({\bf g}), and $\delta A$ ({\bf h}) fluctuations in the IBIS sub-FoVs representative of magnetic and quiet regions. See text for more details. 
	\label{FIG6}}
\end{figure}

Furthermore, we studied the effect of the small magnetic flux concentrations present in the IBIS data on the computed $\lambda$ exponents. As it is known, weak magnetic fields affect the ascending/descending convective motions, thus modifying the typical convection pattern \citep{stein2012} and giving rise to abnormally large granules \citep{tortosa2009,guglielmino2010,guglielmino2018}. 
For our purpose, we evaluated the $\lambda$ exponents for the two sub-FoVs assumed as representative of magnetic and quiet regions, see Figure~\ref{FIG1} (top-left panel) and Figure~\ref{FIG3}. 
Figure~\ref{FIG6} (panel e) shows the $\lambda$ exponents computed from $\delta I_{c}$ fluctuations in the two sub-FoVs. As a reference, the plot also displays the $\lambda$ values obtained from the analysis of the entire, full-resolution IBIS FoV. The average $\lambda$ exponents for the magnetic and quiet sub-FoVs are $-3.9 \pm 0.2$ and $-3.6 \pm 0.1$, respectively, to be compared with $-3.6 \pm 0.1$ obtained from the analysis of the entire IBIS FoV. We note that the values of the $\lambda$ exponent derived from the sub-FoV representative of magnetic regions largely overlap with those derived from the analysis of the quiet and full IBIS FoVs presented above. Figure~\ref{FIG6} (panels f, g, h) display the $\lambda$ exponents computed from $\delta v_{LoS}$, $\delta F$, and $\delta A$ fluctuations in the two sub-FoVs. It can be clearly seen that the $\lambda$ exponents computed from all these quantities from the two sub-FoVs largely overlap. This is even more clear when considering the values of the computed exponents with their uncertainties which are summarized in Table~\ref{TAB3}.

These findings show that the presence of weak magnetic fields only slightly affects the value of the $\lambda$ exponents derived from all the computed fluctuations. For the $\delta I_{c}$ parameter, the interaction between convection and small scale magnetic field concentrations manifests itself with lower values than those derived from quiet regions. 

The aforementioned results refer to the pseudo-Lyapunov exponents $\lambda$ estimated as proposed by \citet{hanslmeier1994}, by using spatial fluctuations of the parameters representative of the studied system
as tracers of either the regular or irregular evolution of the parameters describing the studied system. In order to account for the time dependence in the definition of the pseudo-Lyapunov exponents, we also analyzed the  fluctuations of the computed parameters over time. 

Figure~\ref{FIG7} shows the results obtained from the residual maps of $I_{c}$, $v_{LoS}$, $F$, and $A$ based on the IBIS observations by using Equation~(\ref{EQ3}). The different colors show the results obtained from different values of $j$, which correspond to different time intervals elapsed between the compared observations.

\begin{figure}[H]
	\includegraphics[scale=0.42]{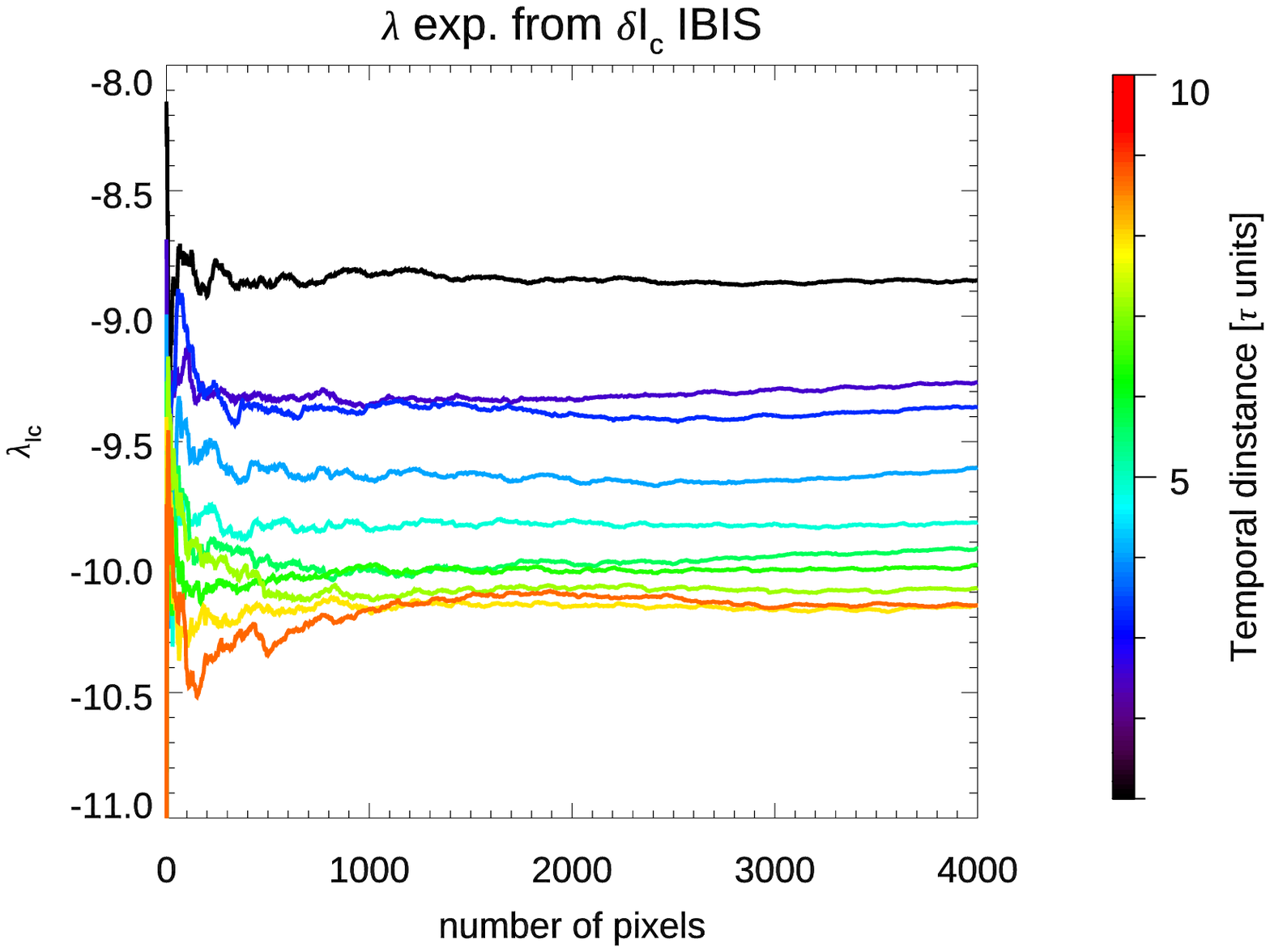}%
	\includegraphics[scale=0.42]{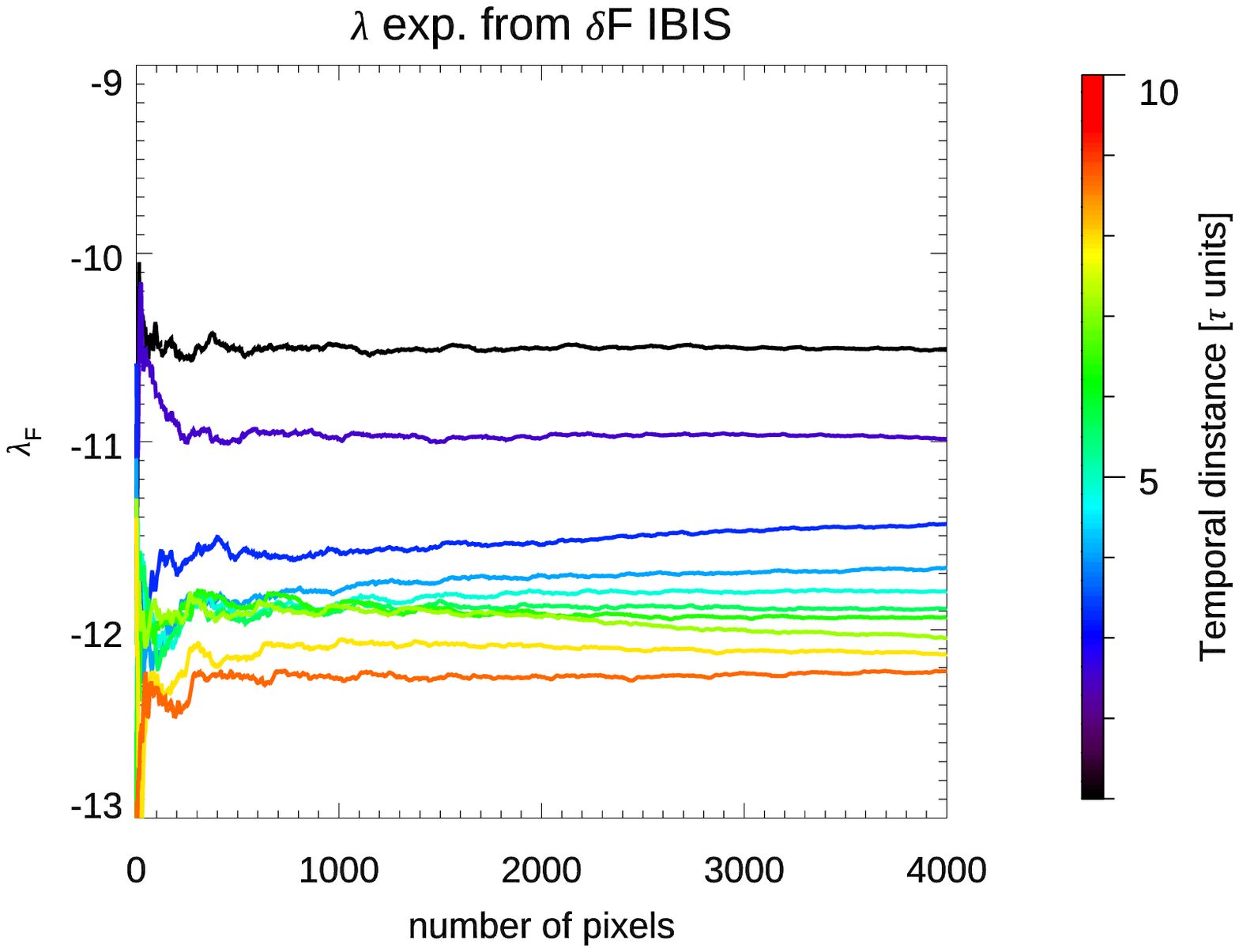}
	\includegraphics[scale=0.42]{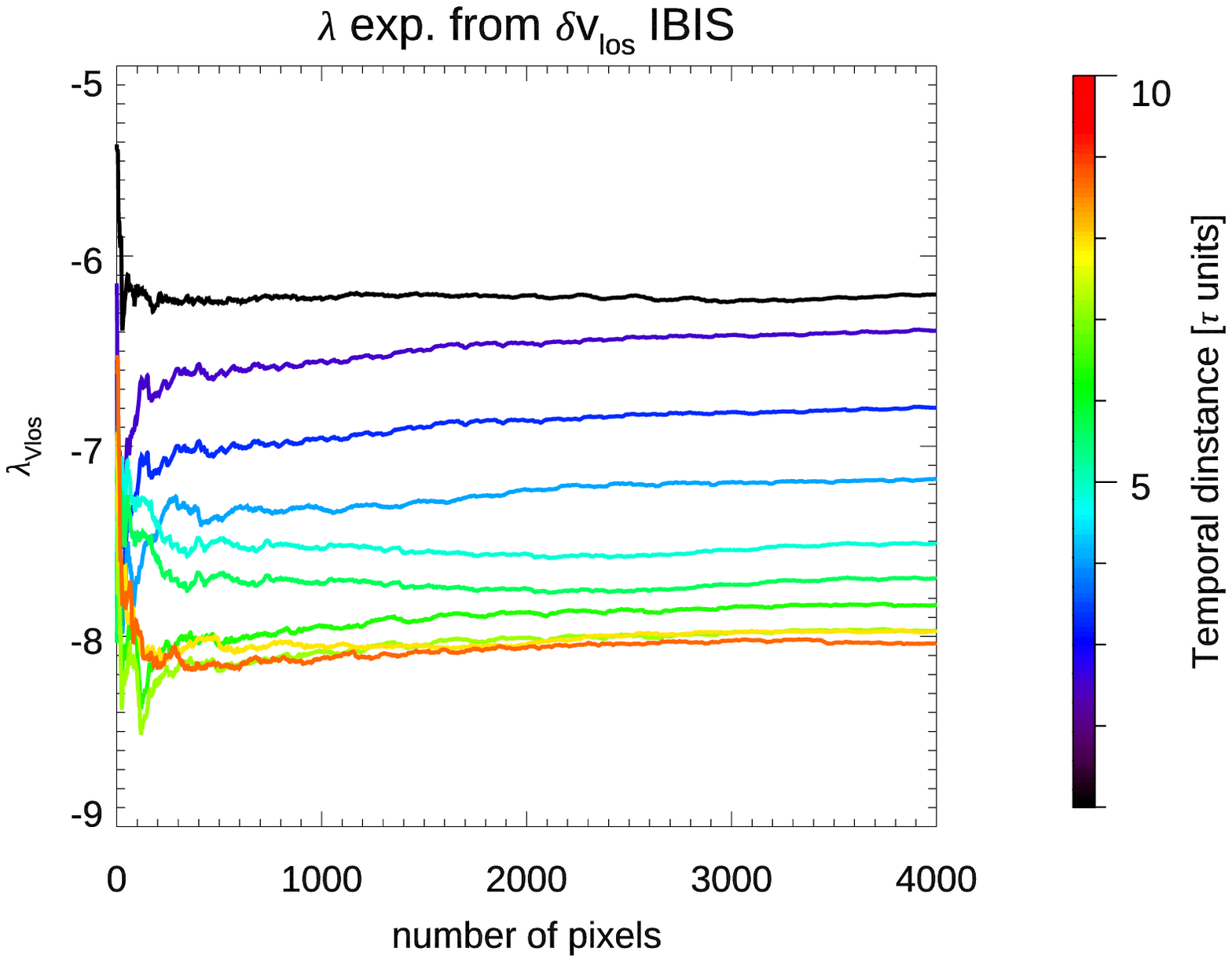}%
	\includegraphics[scale=0.42]{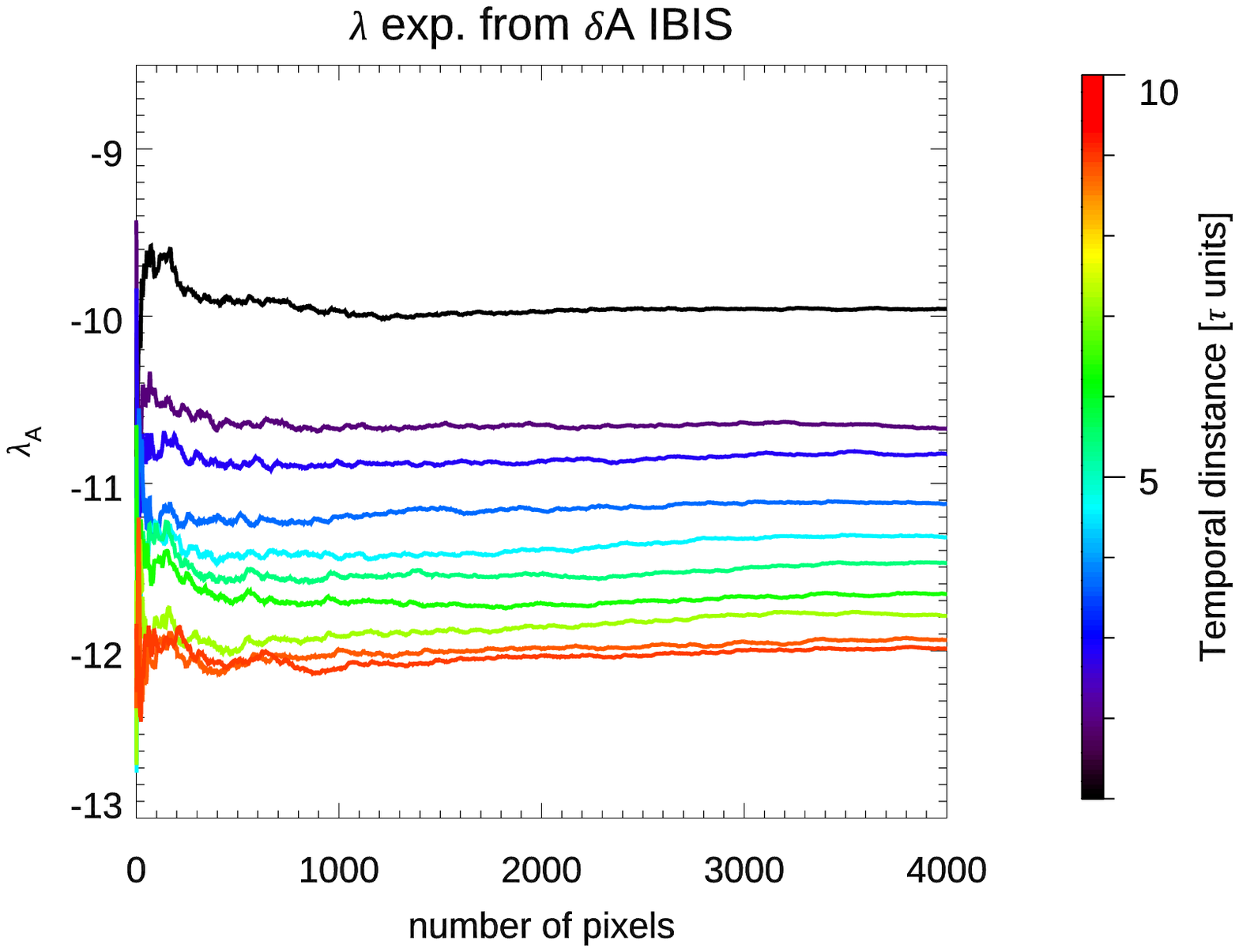}
	\caption{$\lambda$ 
 exponents estimated  from the residuals of $ I_{c}$ (\textbf{top left}), $v_{LoS}$ (\textbf{bottom left}), $FWHM$ (\textbf{top~right}), and $A$ (\textbf{bottom right}) between maps derived from observations taken at different times. The various colors show results achieved by comparing observations taken at increasing regular time intervals in the time series of IBIS data, expressed in multiples of the temporal cadence $\tau$ (48~s). 
	\label{FIG7}}
\end{figure}
We note in Figure~\ref{FIG7} that the exponents derived from the residuals computed over time assume negative values for all the physical quantities estimated from the IBIS data set. Moreover, we report that the estimated exponents clearly tend towards smaller values at the increase in  time elapsed between the compared observations, as for the sampling of a less dissipative regime.

\section{Comments and Conclusions}
\label{sec:final}

In this work, we analyzed the pseudo-Lyapunov exponents $\lambda$ presented by \citet{hanslmeier1994} using state-of-the-art observations taken with the IBIS, CRISP and HMI instruments. Following \citet{hanslmeier1994}, we studied the exponents computed from fluctuations of the continuum intensity $\delta I_{c}$, LoS velocity $\delta v_{LoS}$, full width at half maximum $\delta F$ and asymmetry $ \delta A$ of the spectral lines, which are quantities describing the temperature and velocity of the solar plasma sampled by the observations. The analyzed data sets differ in terms of spatial resolution, spectral sampling, and post facto processing.
We found a dependence of the $\lambda$ exponents on the spatial resolution and other characteristics of the analyzed data. All the computed exponents are negative, in contrast to previous results reported in the literature. 

Our results confirm the dependence of the $\lambda$ exponents on the spatial resolution of the analyzed data reported by \citet{hanslmeier1994}. We showed a  clearer signature of this dependence than previously reported in the literature, which could be suggestive of the detection of less dissipative regimes with increasing spatial resolution. 
In addition, we found higher values of the $\lambda$ exponents obtained from $\delta I_{c}$ fluctuations than in previous analyses, as likely due to more accurate photometry of the data analyzed in our study. We further investigated the exponents in magnetic and quiet regions identified in the IBIS observations, by showing that the values obtained from the two regions coincide within the uncertainty of our estimates.

{Comparing our results with those reported in \citet{hanslmeier1994}, we first noticed that the values computed in our study differ than those previously published.} Our results also exhibit a clear convergence and are less noisy than those in \citet{hanslmeier1994}, especially for the exponents computed from $\delta v_{LoS}$ fluctuations. Conversely, our values are all negative, in contrast to the results in \citet{hanslmeier1994} that show a tendency towards positive values derived from $\delta I_{c}$ and from $\delta A$ fluctuations at some spatial scales through Fourier filtering.  
We inferred that the significant difference between the results obtained in our study, with respect to those by \citet{hanslmeier1994}, may be due to the rather diverse observational techniques employed to obtain the data analyzed in the two compared studies. {In fact, the IBIS and CRISP data were acquired with a spectral sampling performed with Fabry--P\'erot interferometers, while the HMI observations were obtained with a Lyot filter combined with a Michelson interferometer. In contrast, the data used in \citet{hanslmeier1994} are spectrograms acquired with a grating spectrograph. Furthermore}, our data differ from those considered by \citet{hanslmeier1994} in the use of CCD 
sensors instead of photographic plates and of AO systems and of post facto reduction techniques. 
The absence of the AO in the acquisition of the observations analyzed by \citet{hanslmeier1994} and the photographic data they had processed could have induced a smearing of the spatial resolution of the studied observations and of the photometry of their measurements: both these effects can impact the estimated exponents. 
In addition, it is worth noting that \citet{hanslmeier1994} analyzed spectra obtained at a given time, while in our study, we considered measurements obtained from the sampling of a spectral line performed in time. Actually, the different methodology could slightly affect the line shape through which we aimed to infer the properties of the solar convective plasma. {In this respect, it is worth noting that the data used in \citet{hanslmeier1994} are similar for the methodology to the ones acquired within the Solar Optical Telescope (SOT, \citep{tsuneta2008}) aboard the Hinode mission. In addition to the same methodology employed to obtain the data, the SOT/Hinode observations also sample the same spectral line used for the CRISP data analyzed in this study, with images characterized by a slightly lower resolution (approximately 0.3$\arcsec$ for SOT/Hinode with respect to 0.13$\arcsec$ for CRISP). We planned to analyze the SOT/Hinode data in the near future to investigate how the different instrumentation could affect the $\lambda$ exponents determination. Furthermore, we could also analyze the series of HMI observations taken every 45 s, acquired with a different camera than the one used for the data employed in this study, in order to further investigate the exponents derived from time series of observations. Finally, IBIS and CRISP data also allow to derive exponents from line bisector analysis at different line depths that could give more information on the observed processes than those deduced from the simple fitting of the whole spectral line.}

The pseudo-Lyapunov exponents $\lambda$ analyzed by \citet{hanslmeier1994} and in this study are not identical to the Lyapunov exponents $\Lambda$ that, according to the dynamical systems theory, describe the either chaotic or dissipative regime of a system from the evolution of its trajectories in the phase space. The exponents estimated with the method proposed by \citet{hanslmeier1994} assume that the non-linear nature of the solar convection can be represented at any time by the properties of a large sample of convective cells observed simultaneously at different stages. In order to account for the temporal evolution of system entering the definition of the Lyapunov exponents, we also analyzed the variation of the physical quantities used to represent the solar convection at any point over time.
We showed that all the exponents estimated from the residuals of physical quantities analyzed along time assume negative values, which are typical of a dissipative~regime.

Current state-of-the-art observations as those analyzed here allow measuring the properties of solar plasma accurately at spatial scales as small as 100~km at the Sun's surface. However, the resolution and sensitivity of the observations analyzed in our study, as well as both the methods employed to estimate the Lyapunov exponents, have not allowed us to detect any hint of the chaotic regime of solar convection. This regime can only be investigated by resolving even smaller spatial scales than those detected by the observations used in our study, and by using different methods than those employed in this study to measure the divergence of the solar convection in the phase space. The next generation 4 m-class ground-based DKIST and European Solar Telescope (EST, \citep{collados2010}), as well the Polarimetric and Helioseismic Imager \citep{solanki2020} onboard the Solar Orbiter mission \citep{muller2020} are expected to provide spectropolarimetric data sets with unprecedented spatial resolution down to 30~km on the solar atmosphere. The observations that will be obtained with these telescopes appear promising for the study of the small-scale physical processes that occur in turbulent regime  in greater detail,  with methods commonly employed in the dynamical systems theory.

\vspace{6pt} 



\authorcontributions{Conceptualization, all; methodology, S.L.G., I.E., G.C.; software, G.V. and M.M.; validation, S.L.G., I.E., G.C.; formal analysis, G.V. and M.M.; investigation, G.V., M.M., S.L.G. and I.E.; resources, F.G. and S.J.; data curation, F.G. and S.J.; writing---original draft preparation, G.V., M.M., S.L.G. and I.E.; writing---review and editing, all; visualization, G.V., M.M., S.L.G. and I.E.; supervision, S.L.G., I.E., G.C.; project administration, S.L.G.; funding acquisition, I.E. All authors have read and agreed to the published version of the manuscript.}

\funding{This research received funding from the European Union's Horizon 2020 Research and Innovation program under grant agreements No 824135 (SOLARNET) and No 729500 (PRE-EST). This work was supported by the Italian MIUR-PRIN grant 2017 ``Circumterrestrial Environment: Impact of Sun--Earth Interaction'' and by the Istituto Nazionale di Astrofisica. S.J. acknowledges support from the European Research Council under the European Union Horizon 2020 research and innovation program (grant agreement No. 682462) and from the Research Council of Norway through its Centres of Excellence scheme (project No. 262622).}

\institutionalreview{Not applicable. 
}

\informedconsent{Not applicable. 

}

\dataavailability{The IBIS data used in this study can be downloaded from the IBIS-A archive \url{http://ibis.oa-roma.inaf.it/IBISA/database/}.} 

\acknowledgments{This article honors the memory of Anastasios Nesis, who passed away while this paper was being written, remembering his pioneering research on solar convection. {The authors thank the three anonymous Referees for comments that helped to improve the manuscript.}}

\conflictsofinterest{The authors declare no conflict of interest.} 



\abbreviations{The following abbreviations are used in this manuscript:\\

\noindent 
\begin{tabular}{@{}ll}
Ra & Rayleigh number\\
IBIS & Interferometric Bidimensional Spectrometer\\
DST & Dunn Solar Telescope\\
CRISP & Crisp Imaging SpectroPolarimeter\\
SST & Swedish 1 m Solar Telescope\\
HMI & Helioseismic and Magnetic Imager\\
SDO & Solar Dynamic Observatory\\
SHARPs & Space-Weather HMI Active Region Patches\\
FoV & Field-of-view\\
MOMFBD & Multi-Object Multi-Frame Blind Deconvolution\\
AO & Adaptive optics\\
MHD & Magnetohydrodynamic\\
\end{tabular}}

%
%

\end{paracol}
\reftitle{References}

\end{document}